\documentclass[aps,pra,amsmath,superscriptaddress,twocolumn,longbibliography,nofootinbib]{revtex4-1}

\usepackage{graphicx}
\usepackage{amsmath}
\usepackage{amsfonts}
\usepackage{amssymb}
\usepackage{mathrsfs}
\usepackage{braket}
\usepackage{bm}
\usepackage{multirow}
\usepackage{color}
\usepackage[normalem]{ulem}
\usepackage{mathrsfs}
\usepackage{mathtools}
\usepackage{float}
\usepackage{enumitem}

\makeatletter

\newcommand{\Rmnum}[1]{\expandafter\@slowromancap\romannumeral #1@}
\makeatother

\def\ket#1{|\, #1\,\rangle}
\def\bra#1{\langle\,#1\,|}
\def\braket#1#2{\langle\,#1\,|\,#2\,\rangle}
\def\tr{{\rm tr}\,}

\def \sx{\hat\sigma^{\rm x}}
\def \sz{\hat\sigma^{\rm z}}
\def \sy{\hat\sigma^{\rm y}}

\def\P{\widehat P}
\def\PT{{\widehat P}^\top}
\def\S{\widehat {\cal S}}
\def\openone{{1\!\!1}}

\begin{document}

\title{Quantum statistical mechanics of encryption:\\
  reaching the speed limit of classical block ciphers}

\author{Claudio Chamon}
\affiliation{Physics Department, Boston University, Boston, Massachusetts 02215, USA}

\author{Eduardo R. Mucciolo}
\affiliation{Department of Physics, University of Central Florida, Orlando, Florida 32816, USA}

\author{Andrei E. Ruckenstein}
\affiliation{Physics Department, Boston University, Boston, Massachusetts 02215, USA}

\date{\today}

\begin{abstract}
  We cast encryption via classical block ciphers in terms of operator
  spreading in a dual space of Pauli strings, a formulation which
  allows us to characterize classical ciphers by using tools well
  known in the analysis of quantum many-body systems. We connect
  plaintext and ciphertext attacks to out-of-time order correlators
  (OTOCs) and quantify the quality of ciphers using measures of
  delocalization in string space such as participation ratios and
  corresponding entropies obtained from the wave function amplitudes
  in string space. In particular, we show that in Feistel ciphers the
  entropy saturates its bound to exponential precision for ciphers
  with 4 or more rounds, consistent with the classic Luby-Rackoff
  result that it takes these many rounds to generate strong
  pseudorandom permutations. The saturation of the string-space
  information entropy is accompanied by the vanishing of
  OTOCs. Together these signal irreversibility and chaos, which we
  take to be the defining properties of good classical ciphers. More
  precisely, we define a good cipher by requiring that the OTOCs
  vanish to exponential precision and that the string entropies
  saturate to the values associated with a random permutation, which
  are computed explicitly in the paper. In turn, these criteria imply
  that the cipher cannot be distinguished from a pseudorandom
  permutation with a polynomial number of queries. We argue that the
  conditions on both OTOCs and string entropies can be satisfied by
  $n$-bit block ciphers implemented via random reversible circuits
  with ${\cal O}(n \log n)$ gates. This paper focuses on a
  tree-structured cipher composed of layers of $n/3$ 3-bit gates, for
  which a ``key'' specifies uniquely the sequence of gates that
  comprise the circuit. We show that in order to reach this ``speed
  limit'' one must employ a three-stage circuit consisting of a
  nonlinear stage implemented by layers of nonlinear gates that
  proliferate the number of strings, flanked by two linear stages,
  each deploying layers of a special set of linear ``inflationary''
  gates that accelerate the growth of small individual strings.  The
  close formal correspondence to quantum scramblers established in
  this work leads us to suggest that this three-stage construction is
  also required in order to scramble quantum states to similar
  precision and with circuits of similar size. A shallow,
  ${\cal O}(\log n)$-depth cipher of the type described here can be
  used in constructing a polynomial-overhead scheme for computation on
  encrypted data proposed in another publication as an alternative to
  Homomorphic Encryption.
  \end{abstract}

\maketitle

\section{Introduction}
\label{sec:intro}

A block cipher encrypts a plaintext message, broken up into a series
of blocks of bits, by mapping it into ciphertext blocks of the same
size~\cite{Goldwasser-Bellare}. The development of algorithms that
implement ``good'' block ciphers usually involves constructing
pseudorandom permutations through an iterative process which scrambles
the initial plaintext. Notable examples are: (1) Feistel
ciphers~\cite{Goldwasser-Bellare,Luby-Rackoff}, which use pseudorandom
functions to build pseudorandom permutations through multiple rounds
of shuffles and toggles of bit-registers; and (2) random compositions
of small permutations on, at a minimum, 3 bits at a
time~\cite{Coppersmith-and-Grossman, mixwell, mixbetter}. In this
paper we use the latter framework of random classical circuits built
from universal 3-bit gates to explore classical ciphers from a new
point of view. In particular, we formulate plaintext or ciphertext
attacks, which can be cast as combinations of flipping and/or
measuring strings of bits, in terms of out-of-time-order correlators
(OTOCs) of string operators representing the attacks. In our specific
context, the security of a block cipher translates into the
exponential decay of OTOCs as a function of ``computational'' time, a
behavior which in quantum systems signals the approach to a chaotic
state~\cite{Larkin1969,KitaevOTOC,Shenker2014,Shenker2014b,Maldacena2016,
  Aleiner2016,Stanford2016}.

The principal conclusion of this paper is that one can build classical
block ciphers on $n$ bits, secure to polynomial attacks, with as few
as ${\cal O}(n \log n)$ gates in circuits of depth ${\cal O}(\log
n)$. We believe that this result, as well as the physics-inspired
approach proposed in this paper, should have important implications to
many cryptographic applications, especially those requiring fast rates
of encryption when dealing with large data sets. More importantly,
shallow ciphers of this type are the basis for a proposed
Encrypted-Operator Computing (EOC) scheme that allows one to carry out
computations directly on encrypted data via encrypted operators (or
circuits), an alternative to Homomorphic Encryption~\cite{HE_PNAS2015}
presented in Ref.~\cite{EOC}.

The important conceptual element of this paper is to map strings of
Pauli operators describing classical attacks into a dual quantum
mechanical space of strings in which the evolution of string operators
is translated into evolution in a Hilbert space of strings. Within
this string picture, the computation implemented by universal
classical gates evolves an initial string into a {\it superposition}
of string states. The special feature which makes the quantum
mechanical analogy non-trivial is the presence of nonlinear classical
gates $g$ in bit space [i.e., those for which $g(x\oplus y)\ne
  g(x)\oplus g(y)\oplus c$, for constant $c$]. It is this
nonlinearity that leads to the proliferation of string components
making up the quantum superposition, by contrast with linear gates,
which can only change the state of a single string. Thus, as a result
of evolution via nonlinear gates, the string wave function spreads
over the full Hilbert space of strings, reaching an asymptotic state
that cannot be distinguished from a random wave function drawn from an
ensemble that respects all the symmetries of the system. It is the
closeness to this asymptotic state that measures the quality of the
classical cipher, and it is the evolution towards this state that
defines the speed of the scrambling process.

Here we quantify the delocalization of the wave function in string
space in terms of generalized inverse participation
ratios~\cite{Wegner1980}, and corresponding entropies. In the
asymptotic state, these entropies reach their maximum values, which
include a universal correction of order 1 that only depends on the
symmetries of the wave functions, as we illustrate through a
comparison between classical circuits built from 3-bit permutation
gates in $S_8$ and quantum circuits of 2-qubit gates in either $U(4)$
or $O(4)$. The discrete or continuous, real or complex nature of the
wave function in string space is preserved under evolution, defining
three distinct symmetry classes: permutation, orthogonal, and
unitary. Moreover, the asymptotic-state OTOCs vanish, which together
with the maximum entropy reflect the chaotic and irreversible nature
of the evolution.

We view the residual entropy -- the difference from the maximum
entropy -- as the measure of how much information can be extracted by
an adversary. We illustrate this principle by computing the entropies
for a Feistel cipher as function of the number of rounds. We show
that: (a) for 1 or 2 rounds the entropy differs by an extensive amount
from the maximum; (b) for 4 or more rounds the entropy reaches its
maximum up to exponentially-small finite-size corrections; and (c) for
3 rounds the entropy reaches its extensive maximum but with an order 1
deficit with respect to the universal (order 1) correction. These
results are consistent with the conclusions of the classic
Luby-Rackoff work~\cite{Luby-Rackoff}, and reflect the fact that a
Feistel cipher with 4 or more rounds yields a strong-pseudoramdom
permutation, while a 3-round cipher is ``marginal'' in that a combined
3-query plaintext/ciphertext attack (that exploits the regular
structure of the alternating left/right rounds of the Feistel cipher)
can distinguish the resulting permutation from a strong-pseudoramdom
one~\cite{Patarin}.

Arguably, the main contribution of this paper is to determine, for a
cipher built via random reversible classical circuits, the minimum
number of 3-bit gates needed to reach the asymptotic random
multi-string wave function. Physically, the evolution to this
asymptotic state occurs through quantum diffusion in string space,
controlled by transition amplitudes between string states induced by
the action of gates in bit space. We study the statistics of these
transition amplitudes and their associated transition probabilities
for the case of interest, the 3-bit permutation gates, and compare it
to the cases of 2-qubit quantum orthogonal and unitary gates. As shown
explicitly in this paper, the essential feature in all three symmetry
classes is that the evolution via universal gates leads to string
proliferation due to non-zero matrix elements between an initial
string state and multiple final states.

As already mentioned above, we find that the minimum size circuit
leading to the vanishing of the OTOCs to exponential precision and the
saturation of the entropy scales as ${\cal O}(n\log n)$. In the
context of the cipher, this implies that the action of a random
circuit of this size cannot be distinguished from a pseudorandom
permutation with a polynomial number of queries. Establishing this
result requires eliminating a bottleneck associated with the
probability that small initial strings do not grow sufficiently
fast. This bottleneck occurs because the evolution through generic
gates results in a non-zero stay-probability, $p$, for size (or
weight) 1 substrings; which, in turn, translates into a tail in the
distribution of string sizes that scales as $p^\ell$ after the
application of $\ell$ layers of $n/3$ non-overlapping 3-bit
gates. These tails lead to an undesirable polynomial in $n$ decay of
the OTOCs for $\ell\sim \log n$. We eliminate these tails by
structuring the circuit so as to separate two distinct processes: (a)
the extension of a small weight string in the initial state to a
string of macroscopic weight; and (b) the splitting of the resulting
macroscopic string into a superposition of exponentially many string
states, a process necessary for the decay of the residual entropy and
the OTOCs. We identify specific subsets of gates of $S_8$ that
separately implement these processes, which we refer to as
``inflation'' and ``proliferation'', respectively. Inflation is
implemented via circuits built by drawing from a set of 144 special
linear gates (out of the $8!$ 3-bit permutation gates) for which the
stay-probability for weight 1 substrings vanishes. In bit space, these
gates flip two or more bits at the output when a single bit is flipped
at the input. String proliferation is implemented using a subset of
$S_8$ with 10752 ``super-nonlinear'' 3-bit gates that maximize entropy
production.

We use the special features of these gate sets to build a three-stage
cipher, with a circuit of super-nonlinear gates bookended by two
circuits of inflationary gates. Each of these circuits is structured
in layers of 3-bit gates that cover all $n$ bitlines (we assume for
simplicity that $n=3^q$, for an integer $q$). The wiring of the
circuit -- the choice of triplets of bits that are acted upon by 3-bit
gates in each and every layer -- follows a hierarchical (tree)
structure. This choice allows us to carry out our calculations
analytically, and most importantly, it accelerates scrambling as it
mimics a system in infinite spatial dimensions. We show that this
three-stage cipher, with ${\cal O}(\log n)$ layers of gates in each
stage, leads to the exponential vanishing of OTOCs and the saturation
of the entropy to the value associated with a random permutation. We
thus claim that the action of our three-stage cipher of
${\cal O}(n\log n)$ gates cannot be distinguished from a random
permutation with a polynomial number of queries.

We should stress that the mapping to string space highlights a
connection between classical ciphers and the problem of scrambling by
random quantum
circuits~\cite{Oliveira2007,Harrow2009,Brown2013scrambling,Brown2015,
  Brando2016,Harrow2018approximate}. This problem has engendered a
great deal of interest in the context of the recent demonstration of
quantum supremacy~\cite{Google}, as well as in studies of information
processing in black holes, which are conjectured to be the fastest
scramblers in
nature~\cite{Hayden2007,Sekino2008,Shenker2014,Shenker2014b,
  Maldacena2016,Stanford2016}. In the case of random quantum circuits,
resolving the bottleneck associated with the growth of small strings
that we mentioned above, requires circuits of sizes ${\cal
  O}(n\log^2n)$~\cite{Brown2013scrambling,Brown2015,
  Harrow2018approximate}. Our work suggests that, as in the classical
cipher, one can further reduce the circuit size to ${\cal O}(n\log n)$
gates in the quantum case by deploying the three-stage construction
described here. It remains to be determined whether the inflationary
period can be implemented via the unstructured (i.e., random)
placement of 2-qubit gates or it requires using our 144 linear 3-bit
gates or their circuit-equivalent built from the correlated placement
of 2-qubit CNOT gates.

Finally, we expect that ciphers based on random circuits are immune to
quantum attacks. Known vulnerabilities to such attacks arise as a
result of periodicities induced by design regularities, such as the
right/left structure of rounds in Feistel
ciphers~\cite{quantum-attacks}, regularities that are absent in both
classical and quantum random circuits.

The plan of the paper is as follows: in Sec.~\ref{sec:prelim} we
introduce notation and formulate classical ciphers and general attacks
in terms of OTOCs, the vanishing of which establish a criterium for
cipher security. With the notation and formal framework in place, we
are in position to summarize our key conceptual contributions in
Sec.~\ref{sec:contributions}, setting up the roadmap for the rest
of the paper. In Sec.~\ref{sec:stringspace} we formulate the dynamics
of the Pauli operators entering the OTOCs as a quantum evolution
problem in a dual string space. In this subsection we also introduce
generalized inverse participation ratios and their associated
entropies for quantifying the delocalization of the wave function in
string space. In Sec.~\ref{sec:equilibrium} we derive the equilibrium
distribution for the string wave function amplitudes for random
permutations, and compare with those obtained for random unitary and
orthogonal transformations, and extract bounds on the entropies for
all three symmetry classes (see
subsection~\ref{sec:quantum-ensambles}). Subsection~\ref{sec:feistel}
focuses on the case of permutations and uses entropies as diagnostics
for the security of Feistel ciphers, providing a direct connection to
the well-known results of Luby and Rackoff~\cite{Luby-Rackoff}. In
subsection~\ref{sec:OTOC} we establish the vanishing of OTOC in the
equilibrium state characterized by independently and identically
distributed string amplitudes, $A_{\beta\alpha}$. We turn to the
dynamics of the approach to the asymptotic equilibrium state in
Sec.~\ref{sec:dynamics}, where we emphasize the common origin of
string spreading in the structure of the string-space transition
matrix elements for the three symmetry classes of gates, and argue for
the same universal scaling of the minimum size circuit and the
equilibration times for classical and quantum circuits. In the same
section we discuss the subtlety connected with long tails in the
distribution of string weights that slow down scrambling by both
classical and quantum random circuits. In
subsections~\ref{sec:inflation} and ~\ref{sec:proliferation} we single
out two distinct processes - string ``inflation'' and string
``proliferation'' - and the respective sets of gates in $S_8$ that
implement them: linear ``inflationary'' gates which grow/inflate
single strings and eliminate the above mentioned long tails in the
distribution of string weights, and super-nonlinear gates which
accelerate the proliferation in the number of strings, and are
responsible for the saturation of string entropies.  These two sets of
gates are deployed in Sec.~\ref{multi-stage} in the construction of a
three-stage cipher - a central element of this work.  In
subsection~\ref{tree-structured-cipher} we introduce a tree-structured
wiring of the three-stage cipher circuit that accelerates scrambling,
and also allows us to establish the main results of the paper
analytically.  Some important properties of the inflationary and
proliferation stages of the tree-structured cipher are discussed in
subsections~\ref{sec:inflation-with-tree} and
~\ref{sec:proliferation-with-tree}, respectively. These are then
brought together in subsection~\ref{sec:sac-recursion}, where we
present recursion relations for the Strict Avalanche Criterium (SAC)
OTOC for arbitrary numbers of layers of gates, for each of the three
stages of the cipher circuit. The subsection illustrates the behavior
of these recursion relations, explains the mechanism for the
exponential decay of the SAC OTOC with $n$, and documents the
agreement with direct numerical simulations of the OTOC. Taken
together, the results of Sec.~\ref{multi-stage} confirm that the
tree-structured three-stage circuits provide an implementation of
classical ciphers secure to polynomial attacks with as few as
${\cal O}(n \ln n)$ gates. The paper ends in
Sec.~\ref{sec:conclusions} with concluding remarks and questions for
future research.

\section{Block ciphers and reversible computation}
\label{sec:prelim}

A block cipher is a permutation, $P$, that acts on the space of binary
states of $n$ bits, i.e., $P$ is an element of the symmetric group
$S_{2^n}$. Permutations can be thought of as reversible classical
computations, which can be encoded in circuits of universal reversible
gates acting on a small number of bits. More precisely, even
permutations can be decomposed into products of small permutations
(elements of $S_8$) acting on 3 bits at a time, as shown by
Coppersmith and Grossman~\cite{Coppersmith-and-Grossman}. (The
realization of odd permutations requires either one additional $n$-bit
gate or one ancilla bit.) These small permutations can be represented
using a set of universal reversible gates, for example NOT, CNOT, and
Toffoli gates~\cite{Fredkin1982}; here we work directly with the small
permutation gates in $S_8$.

\subsection{Notation}

Hereafter we use quantum bracket notation, and represent a permutation
$P$ acting on a binary string $x\in \{0,1\}^n$ as an operator $\P$
acting on a state $\ket{x}\equiv\ket{x_0\,x_1\dots x_{n-1}}$,
\begin{align}
  {\P}\;\ket{x} = \ket{P(x)}
  \;.
\end{align}
The permutation operator $\widehat P$ is unitary and real:
$\P^{-1}=\P^\dagger=\PT$.

In this representation, reading and flipping bits are implemented
using the Pauli operators $\sz_i$ and $\sx_i$, respectively:
\begin{align}
  &\sz_i\;\ket{x} = (-1)^{x_i}\;\ket{x}
  \nonumber \\
  &\sx_i\;\ket{x} = \ket{x_0\,x_1\dots {\bar x}_i\dots x_{n-1}}
  \;,
\label{eq:pauli}
\end{align}
where ${\bar x}_i$ is the negation of $x_i$. We also introduce
$P_i(x)$ to denote the $i$-th bit of $P(x)$, which can be read via
\begin{align}
  \PT\;\sz_i\;\P\;\ket{x}
  &=
  (-1)^{P_i(x)} \,\ket{x}
  \;.
\end{align}
Since the operator $\hat P$ evolves the input state $\ket{x}$ into the
output state $\ket{P(x)}$, it is natural to interpret
\begin{align}
  \sz_i(\tau)\equiv \PT\;\sz_i\;\P
\end{align}
as the Heisenberg evolution of the operator $\sz_i$ in the course of
the computation. Here $\tau$ defines the accumulated ``time'' of the
computation, which counts the number of gates (or layers of gates as
appropriate) of the circuit implementing the permutation $P$. More
generally, we define $ {\widehat O}(\tau)\equiv \PT\;{\widehat O}\;\P
$ as well as $ {\widehat O}(0)\equiv {\widehat O} $ which represent
operators at the output and input ends of the cipher, respectively.

\subsection{Cryptoanalysis via correlation functions}

In the context of the quantum language defined above, a criterium for
``good'' ciphers will be expressed through the behavior of a class of
correlation functions known as out-of-time-order correlators
(OTOCs)~\cite{Larkin1969,KitaevOTOC,Shenker2014,Shenker2014b,Maldacena2016,
  Aleiner2016,Stanford2016}. Following cryptoanalysis, we diagnose the
quality of the ciphers in terms of plaintext and ciphertext attacks
corresponding to arbitrary readouts and flips on both inputs and
outputs.

Consider first a simple example of plaintext attack in which an
adversary probes the sensitivity of the output bit $j$ to flipping an
input bit $i$, as expressed by
\begin{align}
  \label{SAC-def}
  C^{ij}_{\rm SAC}
  &=
  \frac{1}{2^n}\;\sum_x\,(-1)^{P_j(x\oplus c_i)\oplus P_j(x)}
  \;,
\end{align}
with $c_i=2^i$ and the bitwise XOR operation for two $n$-bit strings
$x\oplus c_i$ implementing the flip of the $i$-th bit of $x$. The
Strict Avalanche Criterium (SAC) test~\cite{Feistel,Llyod,Hirose1995}
requires the function $P_j(x\oplus c_i)\oplus P_j(x)$ to be balanced,
i.e., to be 0 or 1 with equal frequency, and thus $C_{\rm SAC}$ to
vanish. For random permutations, $C_{\rm SAC}$ vanishes up to
corrections that are exponentially small in $n$. Notice that
Eq.~(\ref{SAC-def}) requires summing over all $2^n$ initial states
$x$; in a practical attack, only a number of samples $M$ (polynomial
in $n$) is accessible, in which case an adversary cannot resolve
$C_{\rm SAC}$ below a noise level of ${\cal O}(1/\sqrt{M})$.

Within the quantum notation, Eq.~(\ref{SAC-def}) can be cast as a
correlation function,
\begin{align}
  \label{eq:SAC}
  C^{ij}_{\rm SAC}
  &=
  \frac{1}{2^n}\;\sum_x
  \,\bra{x}\;\sx_i(0)\;\sz_j(\tau)\;\sx_i(0)\;\sz_j(\tau)\;\ket{x}
  \nonumber\\
  &\equiv
  \tr\left[
    \rho_\infty\;\sx_i(0)\;\sz_j(\tau)\;\sx_i(0)\;\sz_j(\tau)
    \right]
  \;,
\end{align}
where $\rho_\infty\equiv \openone/2^n$ can be viewed as an infinite
temperature density matrix. This correspondence is evidenced by
following the sequence of operators from right-to-left in the first
line of Eq.~(\ref{eq:SAC}):
\begin{enumerate}

\item $\sz_j(\tau)$ measures the value of the $j$-th bit,
  $(-1)^{P_j(x)}$, at the output and returns the system to the 
  initial input state, $\ket{x}$;

\item $\sx_i(0)$ flips the $i$-th bit of the initial input state,
  $\ket{x}$, into $\ket{x\oplus c_i}$;

\item $\sz_j(\tau)$ measures the $j$-th bit, $(-1)^{P_j(x\oplus
  c_i)}$, at the output and returns the system to the input state
  $\ket{x\oplus c_i}$; and finally,

\item $\sx_i(0)$ flips the $i$-th bit back,
returning the system to the initial state $\ket{x}$.

\end{enumerate}  

\begin{widetext}
As a second more complex example, we consider an attack that
distinguishes a Feistel cipher build via a 3-round Luby-Rackoff
construction from a strong pseudorandom permutation. This cipher is
vulnerable to a classical adaptive chosen plaintext and chosen
ciphertext attack (CPCA)~\cite{Patarin}, involving three queries,
which we translate into the following OTOC:
\begin{align}
  \label{eq:CPCA}
  C^{ij}_{\rm CPCA}
  &=
  \tr\left[
    \rho_\infty\;
    \sx_j(0)\;\sx_i(\tau)
    \;\sz_i(0)\;\sx_i(\tau)
    \;\sz_j(\tau)\;\sx_j(0)
    \;\sz_j(\tau)\;\sz_i(0)
    \;
    \right]
  \nonumber\\
  &=
  (-1)^{\delta_{ij}}\;
  \tr\left[
    \rho_\infty
    \;\left(\sz_i(0)\;\sx_i(\tau)\right)^2
    \;\left(\sz_j(\tau)\;\sx_j(0)\right)^2
    \;
    \right]
  \;.
\end{align}
We note that the 4-round version of this cipher is immune to this
attack, as are ciphers based on sufficiently long random reversible
circuits.

\section{Summary of conceptual contributions}
\label{sec:contributions}

Having introduced the formal framework, we are in position to
outline the four conceptual contributions of this paper. The first
one is to represent any plaintext/ciphertext attack involving multiple
readouts and/or flips of bits at both inputs and outputs in the form
of an OTOC of Pauli string operators
\begin{align}
  \label{eq:string-alpha}
  \S_\alpha
  =
  \prod_{j\in\alpha^{\rm x}}\;\sx_j
  \;
  \prod_{k\in\alpha^{\rm z}}\;\sz_k
  \;,
\end{align}
namely 
\begin{align}
\label{eq:general-OTOC}
  C^{\,\alpha_1,\beta_1\dots\alpha_k,\beta_k}_{\rm CPCA} &= \tr\left[
    \rho_\infty\;
    \S_{\alpha_1}(0)\;\S_{\beta_1}(\tau)\;
    \S_{\alpha_2}(0)\;\S_{\beta_2}(\tau)\;
    \dots\;
    \S_{\alpha_k}(0)\;\S_{\beta_k}(\tau)
    \right] \;.
\end{align}
\end{widetext}
A Pauli string is labeled by the set
$\alpha=(\alpha^{\rm x},\alpha^{\rm z})$ of bit indices present in the
string.  (We choose to place all the $\sx$ operators to the left of
the $\sz$s~\footnote{By adding a phase
  $i^{\alpha^{\rm x}\cdot\alpha^{\rm z}}$ to $\S_\alpha$ -- picking up
  an $i$ each time both a $\sx_j$ and $\sz_j$ appear at the same $j$,
  or basically deploying the $\sy$s as well -- would make the string
  operator Hermitian. Here we prefer the definition
  Eq.~(\ref{eq:string-alpha}) for the applications we consider, and
  work explicitly with both $\S^{\; }_\alpha$ and
  $\S^{\dagger}_\alpha$ when needed.}.)

The second conceptual contribution is to tie the security of the
cipher to the vanishing of OTOCs representing plaintext/ciphertext
attacks. In the quantum case, for both Hamiltonian systems and
evolution via random quantum circuits, the vanishing of OTOCs is
associated with irreversibility and
chaos~\cite{KitaevOTOC,Shenker2014,Shenker2014b,Maldacena2016,
  Aleiner2016,Stanford2016}.

The third contribution is to connect the vanishing of OTOCs to the
delocalization of the dual string-space wave function via the growth
in size and exponential proliferation in number of Pauli strings in
the course of the computation. We quantify this delocalization in
terms of generalized inverse participation ratios~\cite{Wegner1980}
and corresponding entropies, which saturate for a random string wave
function, thus tying together the vanishing of the OTOCs and the
vanishing of the residual entropies for random permutations.

Finally, the fourth contribution is to realize that one must break the
circuit into three stages, separating the string-space
``inflationary'' and ``proliferation'' action of special collections
of gates in $S_8$. It is this three-stage structure that allows us to
construct a cipher with as few as ${\cal O}(n\log n)$ gates for which
OTOCs vanish to exponential precision, and the string entropies
saturate.

\section{Quantum evolution in string space}
\label{sec:stringspace}

The description of OTOCs in terms of strings in
Eq.~\eqref{eq:general-OTOC} brings out the connection to quantum
mechanics that is critical to our analysis: time evolution leads to
superpositions in string space. It is this correspondence that allows
us to unify results in classical and quantum random circuits.

We further note that while classical permutations are the principal
motivation for this work, below we use $\widehat P$ to denote a
general unitary transformation, since in considering the connection to
quantum computation $\widehat P$ will be replaced by a unitary
$\widehat U$ or orthogonal $\widehat O$ transformation. The special
properties of permutation operators will be noted as needed.

The connection to quantum mechanics is made manifest by translating
the string operators $\widehat{\mathcal{S}}_\beta$ and
$\widehat{\mathcal{S}}_\alpha^{\;\dagger}$ defined by
Eq.~\eqref{eq:string-alpha} into bra $\bra\beta$ and ket $\ket\alpha$
states in a dual string space that inherits the inner product
\begin{align}
  \braket{\beta}{\alpha}
  &=
  \frac{1}{2^n}\;
  \tr \left({\widehat S}_\beta^{\; }\;{\widehat S}_\alpha^{\;\dagger}\right)
    =
    \delta_{\beta\alpha}
    \;.
\end{align}
The states resulting from this operator-to-state correspondence should
be viewed as tensor products of $n$ qudits that label 4 possible Pauli
operators at each position in the string. The Heisenberg evolution of
the string operators
\begin{align}
{\widehat S}_\beta(\tau)\equiv \widehat
P^{\dagger}(\tau)\, {\widehat S}_\beta\, \widehat P(\tau) =
\sum_\alpha A_{\beta\alpha}(\tau)\; {\widehat S}_\alpha
\label{eq:time-evol-Sbeta}
\end{align}
corresponds to the evolution in Hilbert space of string states via the
unitary operator $\widehat{{\cal U}}_P$, namely,
\begin{align}
\bra{\beta(\tau)} =
\bra{\beta}\; \widehat{{\cal U}}_P
=
\sum_\alpha A_{\beta\alpha}(\tau)\;\bra{\alpha}
  \;,
\end{align}
where the transition amplitudes between string states $\alpha$ and
$\beta$ are given by
\begin{align}
  A_{\beta\alpha}(\tau)
  &=
  \bra{\beta}\;\widehat{{\cal U}}_P \ket{\alpha}
  \nonumber\\
  &= 
  \frac{1}{2^n}\;
  \tr \left(\widehat P^{\dagger} \; {\widehat S}_\beta^{\;}\; \widehat P
  \; {\widehat S}_\alpha^{\;\dagger}\right)
  \;.
\end{align}
We note that if one breaks the operators $\widehat P$ and $\widehat
P^\dagger$ above into gates, these appear sequentially bookending the
operator ${\widehat S}_\beta$ in reverse-time order or, using the
cyclic property of the trace, the operator ${\widehat
  S}^{\;\dagger}_\alpha$ in the natural time order. Thus, the
amplitudes $A_{\beta\alpha}(\tau)$ can be viewed as describing either
forward propagation from $\alpha$ to $\beta$ or backward propagation
from $\beta$ to $\alpha$. Moreover $A_{\beta\alpha}(\tau)$ satisfies
the normalization conditions
\begin{align}
  \label{eq:norm}
  \sum_\beta |A_{\beta\alpha}(\tau)|^{\;2}=
  \sum_\alpha |A_{\beta\alpha}(\tau)|^{\;2}=1
  \;,
\end{align}
which follow from the unitarity of $\widehat{{\cal U}}_P$, itself a
consequence of preservation of the norm under time evolution:
\begin{align}
  \bra{\beta}\;\widehat{{\cal U}}_P
  \; \widehat{{\cal U}}^{\;\dagger}_P \ket{\alpha}
  &=\braket{\beta(\tau)}{\alpha(\tau)}
\nonumber\\
  &=\frac{1}{2^n}\; \tr \left(\widehat
P^{\dagger} \; {\widehat S}_\beta^{\;}\; \widehat P \; \widehat
P^{\dagger} \; {\widehat S}_\alpha^{\;\dagger}\; \widehat P \right)
\nonumber\\
&= \braket{\beta}{\alpha}
\;.
\end{align}


Using the string amplitudes $A_{\alpha\beta}$ we can re-express the
OTOC in Eq.~\eqref{eq:general-OTOC}:
\begin{widetext}
\begin{align}
\label{eq:general-OTOC2}
C^{\,\alpha_1,\beta_1\dots\alpha_k,\beta_k}_{\rm CPCA}
&=
\sum_{\gamma_1,\dots,\gamma_k}
A_{\beta_1\gamma_1}(\tau)\;A_{\beta_2\gamma_2}(\tau)\;\dots\;A_{\beta_k\gamma_k}(\tau)\;
\tr\left[
  \rho_\infty\;
  \S_{\alpha_1}\;\S_{\gamma_1}\;
  \S_{\alpha_2}\;\S_{\gamma_2}\;
  \dots\;
  \S_{\alpha_k}\;\S_{\gamma_k}
  \right]
\nonumber\\
&=
\sum_{\gamma_1,\dots,\gamma_k}
A_{\beta_1\gamma_1}(\tau)\;A_{\beta_2\gamma_2}(\tau)\;\dots\;A_{\beta_k\gamma_k}(\tau)\;
\;
(-1)^{\sum_{i\le j} \alpha^{\rm z}_i\cdot\gamma^{\rm x}_j}\;
(-1)^{\sum_{i<j} \gamma^{\rm z}_i\cdot\alpha^{\rm x}_j}\;
(-1)^{\sum_{i< j} \alpha^{\rm z}_i\cdot\alpha^{\rm x}_j}\;
(-1)^{\sum_{i<j} \gamma^{\rm z}_i\cdot\gamma^{\rm x}_j}\;
\nonumber\\
&\hspace{4cm}\times
\delta_{
  \alpha^{\rm x}_1\oplus\dots\oplus\alpha^{\rm x}_k,
  \gamma^{\rm x}_1\oplus\dots\oplus\gamma^{\rm x}_k}\;\;\;
\delta_{
  \alpha^{\rm z}_1\oplus\dots\oplus\alpha^{\rm z}_k,
  \gamma^{\rm z}_1\oplus\dots\oplus\gamma^{\rm z}_k}
\;.
\end{align}
\end{widetext}
where the dot product is defined as
$a\cdot b\equiv a_1b_1+\dots+ a_nb_n$. An important quantity that will
be used throughout the paper is the string weight, defined as
$w(\alpha)\equiv \sum_i (\alpha^{\rm x}_i \lor \alpha^{\rm z}_i)$,
which measures the number of non-identity Pauli operators that compose
the string. (Each $\alpha^{\rm x}_i \lor \alpha^{\rm z}_i$ is 1 if at
location $i$ the string has an $\sx_i$, an $\sz_i$, or both operators,
and is 0 otherwise.)

Note that the string operators in Eq.~\eqref{eq:string-alpha} were
conveniently defined as product of real operators, using $\sx$ and
$\sz$s, to avoid factors of $i$ when making the connection with the
cipher attacks. The effect of $\sy$s appears when $\sx$ and $\sz$
operators overlap in the string ${\widehat S}_\alpha$; we count the
number of such overlaps by
$n_y(\alpha)\equiv {\alpha^{\rm x}\cdot\alpha^{\rm z}}$. These strings
are not self-conjugate and satisfy
\begin{align}
  {\widehat S}_\alpha^{\;\dagger}
  = (-1)^{\alpha^{\rm x}\cdot\alpha^{\rm z}}
  \;{\widehat S}_\alpha^{\;}
  \;,
\end{align}
which, in turn, leads to the following symmetry relation, preserved
throughout the evolution:
\begin{align}
  \label{eq:symmetry_parity}
  A^*_{\beta\alpha}(\tau)
  = (-1)^{\beta^{\rm x}\cdot\beta^{\rm z}}\;(-1)^{\alpha^{\rm x}\cdot\alpha^{\rm z}}
  \;A_{\beta\alpha}(\tau)
  \;.
\end{align}

For permutations, the amplitudes are real, and this symmetry implies
that $A_{\beta\alpha}(\tau)$ is non-zero if and only if
$n_y(\alpha)=n_y(\beta)\!\!\mod 2$. In terms of the permutations
$P(x)$ on the bit strings $x$, the amplitudes take values in a
discrete set, as can be seen from the explicit formula:
\begin{widetext}
\begin{align}
  \label{eq:explicit_A_from_tr}
  A_{\beta\alpha}(\tau)
  &=
  \frac{1}{2^n}\;
  \sum_x\;
  \bra{x}\;\widehat P^{\dagger} \; {\widehat S}_\beta^{\;}\; \widehat P\;{\widehat S}_\alpha^{\;\dagger}\;
  \ket{x}
  =
  \frac{1}{2^n}\;
  \sum_x\;
  \bra{x}\;\widehat P\;{\widehat S}_\alpha^{\;\dagger}\;\widehat P^{\dagger} \; {\widehat S}_\beta^{\;}\; 
  \ket{x}
  \nonumber\\
  &=
  \frac{1}{2^n}\;
  \sum_x\;
  (-1)^{\alpha^{\rm z}\cdot P^{-1}(x)}\;
  \braket{P^{-1}(x)\oplus \alpha^{\rm x}}{P^{-1}(x\oplus\beta^{\rm x})}\;(-1)^{\beta^{\rm z}\cdot x}
  \nonumber\\
  &=
  \frac{1}{2^n}\;
  \sum_{x}
  \;(-1)^{\beta^{\rm z}\cdot x}
  \;(-1)^{\alpha^{\rm z}\cdot P^{-1}(x)}\;
  \;\delta_{\beta^{\rm x} ,\, x\oplus P(P^{-1}(x)\oplus\alpha^{\rm x})}
  \;.
\end{align}
\end{widetext}

For orthogonal transformations, $\widehat O$, the amplitudes
$A_{\beta\alpha}(\tau)$ are real but continuous, and
Eq.~\eqref{eq:symmetry_parity} implies that $A_{\beta\alpha}(\tau)$ is
non-zero if and only if $n_y(\alpha)=n_y(\beta)\!\!\mod 2$, just as
for permutations. For unitary transformations, $\widehat U$, the
amplitudes are also continuous and Eq.~\eqref{eq:symmetry_parity}
implies two cases according to the relative parity of $n_y(\alpha)$
and $n_y(\beta)$: (a) for $n_y(\alpha)=n_y(\beta)\!\!\mod 2$, the
amplitudes $A_{\beta\alpha}(\tau)$ are real; and (b) for
$n_y(\alpha)\ne n_y(\beta)\!\!\mod 2$, the amplitudes
$A_{\beta\alpha}(\tau)$ are purely imaginary.

We note that our discussion also applies for backpropagation via the
inverse operators (e.g., the inverse permutation $P^{-1}$) by
exchanging $\widehat P$ and $\widehat P^\dagger$ through a
``time-reversal'' transformation:
\begin{align}
  \label{eq:TRS}
  A_{\alpha\beta}(-\tau)
  &\equiv
  \bra{\alpha}\;\widehat{{\cal U}}_{P^{-1}} \ket{\beta}
  \nonumber\\
  &= 
  \frac{1}{2^n}\;
  \tr \left(\widehat P \; {\widehat S}_\alpha^{\;}\; \widehat P^{\dagger}
  \; {\widehat S}_\beta^{\;\dagger}\right)
  \nonumber\\
  &=
  A^*_{\beta\alpha}(\tau)
  \;.
\end{align}

Given the operator-to-state correspondence defined above, the main
results of the paper will follow from the behavior of the string
amplitudes $A_{\beta\alpha}(\tau)$ as functions of $\tau$. These wave
function amplitudes will allow us to quantify (a) the delocalization
in string space in analogy with measures of (de)localization in
quantum systems, and (b) the time, $\tau_{\rm scramble}$, needed for
the amplitudes to approach those of a maximally scrambled state. In
particular, this random state is the equilibrium state with maximum
entropy and vanishing OTOCs. The connection between attacks and OTOCs
implies that the random classical circuit leads to a secure cipher
after time $\tau_{\rm scramble}$.

For a fixed initial string state $\alpha$, the amplitudes
$A_{\beta\alpha}$ represent a wave function in a $d=4^n$ dimensional
Hilbert space. Delocalization in this space can be quantified via the
generalized inverse participation ratios and their associated
entropies that are defined, respectively, by
\begin{align}
  {\cal P}_q
  &=
  \sum_\beta |A_{\beta\alpha}(\tau)|^{\;2q}
  \;,
\end{align}
and
\begin{align}
  S_q = \frac{1}{1-q}\;\ln{\cal P}_q
  \;.
  \label{eq:entropies-def}
\end{align}
These provide concrete and intuitive measures of information
scrambling, reflected in the proliferation of strings and
delocalization in string space. In particular: ${\cal P}_{q\to 0_+}$
measures the number of non-zero amplitudes $A_{\beta\alpha}$, thus
counting the number of strings,
\begin{align}
  {\cal N}_s={\cal P}_{q\to 0_+}
  \;;
\end{align}
$S_{q\to 1}$ gives the information entropy,
\begin{align}
  S = - \sum_\beta |A_{\beta\alpha}(\tau)|^{\;2} \: \ln \, \left(\,|A_{\beta\alpha}(\tau)|^{\;2}\right)
  \;;
\end{align}
and
\begin{align}
  \label{eq:P2}
  {\cal P}_2
  &=
  \sum_\beta |A_{\beta\alpha}(\tau)|^{\;4}
\end{align}
is the inverse participation ratio, familiar from the theory of
localization in quantum systems~\cite{Wegner1980}.

\section{Equilibrium}
\label{sec:equilibrium}

Before we focus on the dynamics of delocalization in string space and
the approach to equilibrium, we discuss the asymptotic equilibrium
state. At equilibrium, the string wave functions $A_{\beta\alpha}$
become independently distributed over string space, i.e., over
different initial and final states $\alpha$ and $\beta$. The
independent distributions are however constrained by the
normalization condition, Eq.~\eqref{eq:norm}, the parity condition,
Eq.~\eqref{eq:symmetry_parity}, as well as the symmetry class of the
underlying transformation defining the computation. Below we derive
the explicit probability distributions for the amplitudes
$A_{\beta\alpha}$ for the three symmetry classes associated with
permutations, orthogonal, and unitary transformations.

\subsection{Random permutations}
\label{sec:random_permutations}

We start by considering the equilibrium statistical properties of the
string amplitudes $A_{\beta\alpha}$ for truly random permutations,
with no reference to how they are built. In particular, we compute the
probability distribution for the string amplitudes $A_{\beta\alpha}$,
from which we obtain the generalized inverse participation ratios and
entropies characterizing the equilibrium state.

The form of the probability distributions of the $A_{\beta\alpha}$
amplitudes can be obtained explicitly in the large-$d$ regime as
follows. Starting from Eq.~\eqref{eq:explicit_A_from_tr}, notice that
the values of $\beta^{\rm x}$ that collect non-zero contributions from
the summation over binary states $x$ must satisfy
\begin{align}
  \beta^{\rm x} = x \oplus \tilde P(x)
  \;,
  \label{eq:relation1}
\end{align}
where
\begin{align}
  \tilde P(x)\equiv P(P^{-1}(x)\oplus\alpha^{\rm x})
  \;.
  \label{eq:relation2}
\end{align}
For any permutation $P$, the associated permutation $\tilde P$ is an
involution, i.e., $\tilde P^2=\openone$, for any $\alpha^{\rm x}$. Now
notice that the pair of state values $x$ and $\tilde P(x)$ contribute
to the same $\beta^{\rm x}$ ``box'', since $x \oplus \tilde P(x) =
\tilde P(x) \oplus \tilde P(\tilde P(x))=\tilde P(x)\oplus x$. We thus
partition the set of $N=2^n$ values of $x$ into $N/2$ pairs
$(x_1,x_2), \dots, (x_{N-1}, x_{N})$, where $x_{2j}=\tilde
P(x_{2j-1}),\, j=1,\dots, N/2$. For a random $P$, the number of pairs
that fall into a given $\beta^{\rm x}$ box (due to $x_{2j-1}\oplus
x_{2j}=\beta^{\rm x}$) should be Poisson distributed , i.e., the
probability that $m$ pairs fall into box $\beta^{\rm x}$ is
\begin{align}
p_m=e^{-1/2}\;\frac{1}{m!}\,\frac{1}{2^m}
  \;,
\end{align}
where the Poisson parameter is $1/2$ (since there are $N/2$ pairs for
$N$ boxes).

Next, let us consider the phases, i.e., signs of contributions to
Eq.~\eqref{eq:explicit_A_from_tr}. The phases coming from the two
entries in a pair interfere either constructively if ${\beta^{\rm
    x}\cdot\beta^{\rm z}}={\alpha^{\rm x}\cdot\alpha^{\rm z}}\!\!\mod
2$, or destructively if ${\beta^{\rm x}\cdot\beta^{\rm z}}\ne
{\alpha^{\rm x}\cdot\alpha^{\rm z}}\!\!\mod 2$. The two cases can be
considered together by defining $\eta\equiv {\beta^{\rm
    x}\cdot\beta^{\rm z}}+{\alpha^{\rm x}\cdot\alpha^{\rm z}}\!\!\mod
2$. More precisely, by using the identities $P^{-1}(\tilde
P(x))=P^{-1}(x)\oplus \alpha^{\rm x}$ [see Eq.~\eqref{eq:relation2}]
and $\tilde P(x)\oplus \beta^{\rm x}=x$ [see
  Eq.~\eqref{eq:relation1}], we can relate the phases associated with
the two partners, $\tilde P(x)$ and $x$, in a pair as follows:
\begin{align}
  &\;
  (-1)^{\beta^{\rm z}\cdot \tilde P(x)}\;
  (-1)^{\alpha^{\rm z}\cdot P^{-1}(\tilde P(x))}
  \nonumber\\
  =&\;
  (-1)^{\beta^{\rm z}\cdot \tilde P(x)}\;
  (-1)^{\alpha^{\rm z}\cdot P^{-1}(x)}\;
  (-1)^{\alpha^{\rm z}\cdot \alpha^{\rm x}}\;
  \nonumber\\
  =&\;
  (-1)^{\beta^{\rm z}\cdot \tilde P(x)}\;
  (-1)^{\alpha^{\rm z}\cdot P^{-1}(x)}\;
  (-1)^{\beta^{\rm z}\cdot \beta^{\rm x}}\;
  (-1)^\eta
  \nonumber\\
  =&\;
  (-1)^{\beta^{\rm z}\cdot \,[\tilde P(x)\oplus \beta^{\rm x}]}\;
  (-1)^{\alpha^{\rm z}\cdot P^{-1}(x)}\;
  (-1)^\eta
  \nonumber\\
  =&\;
  (-1)^{\beta^{\rm z}\cdot x}\;
  (-1)^{\alpha^{\rm z}\cdot P^{-1}(x)}\;
  (-1)^\eta
\;,
\end{align}
and thus, as claimed above, the two contributions come either with the
same sign if $\eta$ is even, or with opposite signs if $\eta$ is odd.

\begin{widetext}
The interference described above implies that only half of the values
of $\beta^{\rm z}$ result in a non-zero amplitude, which is computed
as follows. Members of each pair will add $\pm 1$ in phase, and thus
contribute either $+2$ or $-2$ to the amplitude. Let $m_\pm$ denote
the number of pairs falling within a $\beta^{\rm x}$ box
contributing $\pm 2$, respectively. The probability that the amplitude
$A=2r/N$ is
  \begin{align}
  \label{eq:prob_A}
  p(A=2r/N)
  &=
  \sum_{m} e^{-1/2}\;\frac{1}{m!}\,\frac{1}{2^m}
  \sum_{m_++m_-=m} \frac{m!}{m_+!\,m_-!}\;\frac{1}{2^{m_+}}\,\frac{1}{2^{m_-}}
  \;\delta_{r,m_+-m_-}
  \nonumber\\
  &=
  e^{-1/2}\;
  \sum_{m_+,m_-} \frac{1}{m_+!}\,\frac{1}{m_-!}\;\frac{1}{4^{m_+}}\,\frac{1}{4^{m_-}}
  \;\delta_{r,m_+-m_-}
  \nonumber\\
  &=
  e^{-1/2}\;
  \sum_{k} \frac{1}{(|r|+k)!}\,\frac{1}{k!}\;\left(\frac{1}{4}\right)^{|r|+2k}
  \nonumber\\
  &=
  \frac{1}{\sqrt{e}}\;
  I_{|r|}(1/2)
  \;,
\end{align}
\end{widetext}
where $I_{\nu}(z)$ is the modified Bessel function.

To compute the moments, it is useful to introduce the generating
function
\begin{align}
  \tilde p(w)
  &=
  \sum_{r=-\infty}^{\infty} p(A=2r/N)\;e^{iwr}
  \nonumber\\
  &=
  e^{-1/2}\;
  \sum_{m_+,m_-} \frac{1}{m_+!}\,\frac{1}{m_-!}\;\frac{1}{4^{m_+}}\,\frac{1}{4^{m_-}}
  \;e^{iw (m_+-m_-)}
  \nonumber\\
  &=
  e^{-\frac{1}{2}}\;
  e^{\frac{1}{2}\cos w}
  \;.
\end{align}
The corresponding (averaged) generalized participation ratios are given by
\begin{align}
  \overline{{\cal P}_q}
  &=
  d^{1-q}\;\;2^{2q-1}\;\;
  \frac{1}{\sqrt{e}}\;
  \left(-i\frac{\partial}{\partial w}\right)^{2q}
  e^{\frac{1}{2}\cos w} \bigg|_{w=0}
  \;.
\end{align}
The associated equilibrium entropies display an extensive (volume)
contribution and a correction, $\Delta{S^{\rm eq}_q}$, of order 1:
\begin{align}
  \label{eq:equil-S}
  {S^{\rm eq}_q}
  &=
  n \ln 4 - \Delta{S^{\rm eq}_q}
\;,
\end{align}
where
\begin{align}
  \label{eq:delta_S_q}
  \Delta{S^{\rm eq}_q}
  &=
  \frac{1}{q-1}\;\ln\left[
  \frac{1}{2\sqrt{e}}\;
  \left(-2i\frac{\partial}{\partial w}\right)^{2q}
  e^{\frac{1}{2}\cos w} \bigg|_{w=0}
  \right]
\;.
\end{align}
(We note that in calculating the entropies we used annealed averages,
since in the regime of delocalized string wave functions the quenched
and annealed averages should coincide up to exponentially small
corrections.)

For integer values, the $\Delta{S^{\rm eq}_q}$ corrections can be read
off from Eq.~\eqref{eq:delta_S_q}: for example
$\Delta{S^{\rm eq}_2} = \ln 10$,
$\Delta{S^{\rm eq}_3} = \frac{1}{2}\ln 196 = \ln 14$, and
$\Delta{S^{\rm eq}_4} = \frac{1}{3}\ln 6280$. The value
$\Delta{S^{\rm eq}_1}=1.9618961$
is obtained numerically from the probalilities in
Eq.~\eqref{eq:prob_A}. These residual entropies are used in the next
sub-section to diagnose the quality of a Feistel cipher.


\subsection{Application to Feistel ciphers}
\label{sec:feistel}

A Feistel cipher, also referred to as a Luby-Rackoff
cipher~\cite{Goldwasser-Bellare,Luby-Rackoff}, builds a pseudorandom
permutation by using pseudorandom functions. The construction of the
cipher proceeds by first separating $n$ bits ($n$ even) into left
($L$) and right ($R$) registers with $n/2$ bits each. Denoting the
input to the cipher by the pair $(L_0,R_0)$, one then iterates through
rounds $k=1,\dots, r$, according to:
\begin{align}
  L_k &= R_{k-1}\nonumber\\
  R_k &= L_{k-1}\oplus f_k(R_{k-1})
  \;,
  \label{eq:feistel-P}
\end{align}
where $f_k$, $k=1,\dots, r$, are different pseudorandom functions with
$n/2\,$-bit inputs and outputs. The iterations in
Eq.~\eqref{eq:feistel-P} define a permutation $P$ such that $(L_r,
R_r)= P_{}\,(L_0,R_0)$. The inverse permutation, $(L_0,R_0) =
P^{-1}_{}(L_r, R_r)$, is obtained by starting with $(L_r, R_r)$ and
working backwards, for $k=r, \dots, 1$, using the recursion
\begin{align}
  R_{k-1} &= L_{k}\nonumber\\
  L_{k-1} &= R_{k}\oplus f_k(L_{k})
  \;.
  \label{eq:feistel-Pinv}
\end{align}
Luby and Rackoff~\cite{Luby-Rackoff} proved that a 3-round (4-round)
cipher is indistinguishable from a random permutation when
interrogated with a polynomial number of one-sided or plaintext
queries (two-sided or plaintext/ciphertext queries). [An example of a
  plaintext/ciphertext attack on a 3-round Feistel cipher was
  represented as an OTOC in Eq.~\eqref{eq:CPCA}.]

\begin{widetext}

\begin{figure}[ht]
\centering
\includegraphics[width=0.45\textwidth]{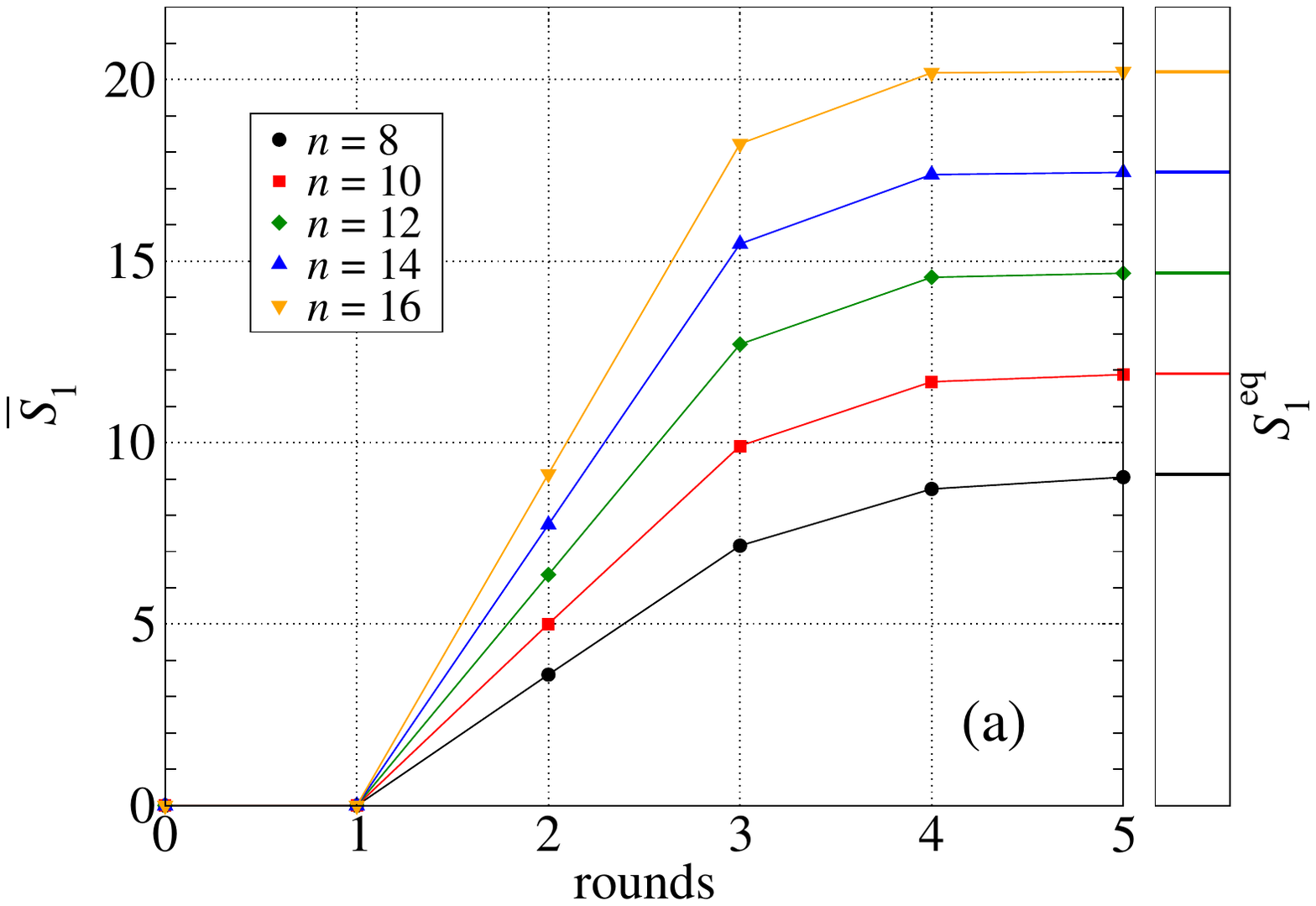}
\includegraphics[width=0.45\textwidth]{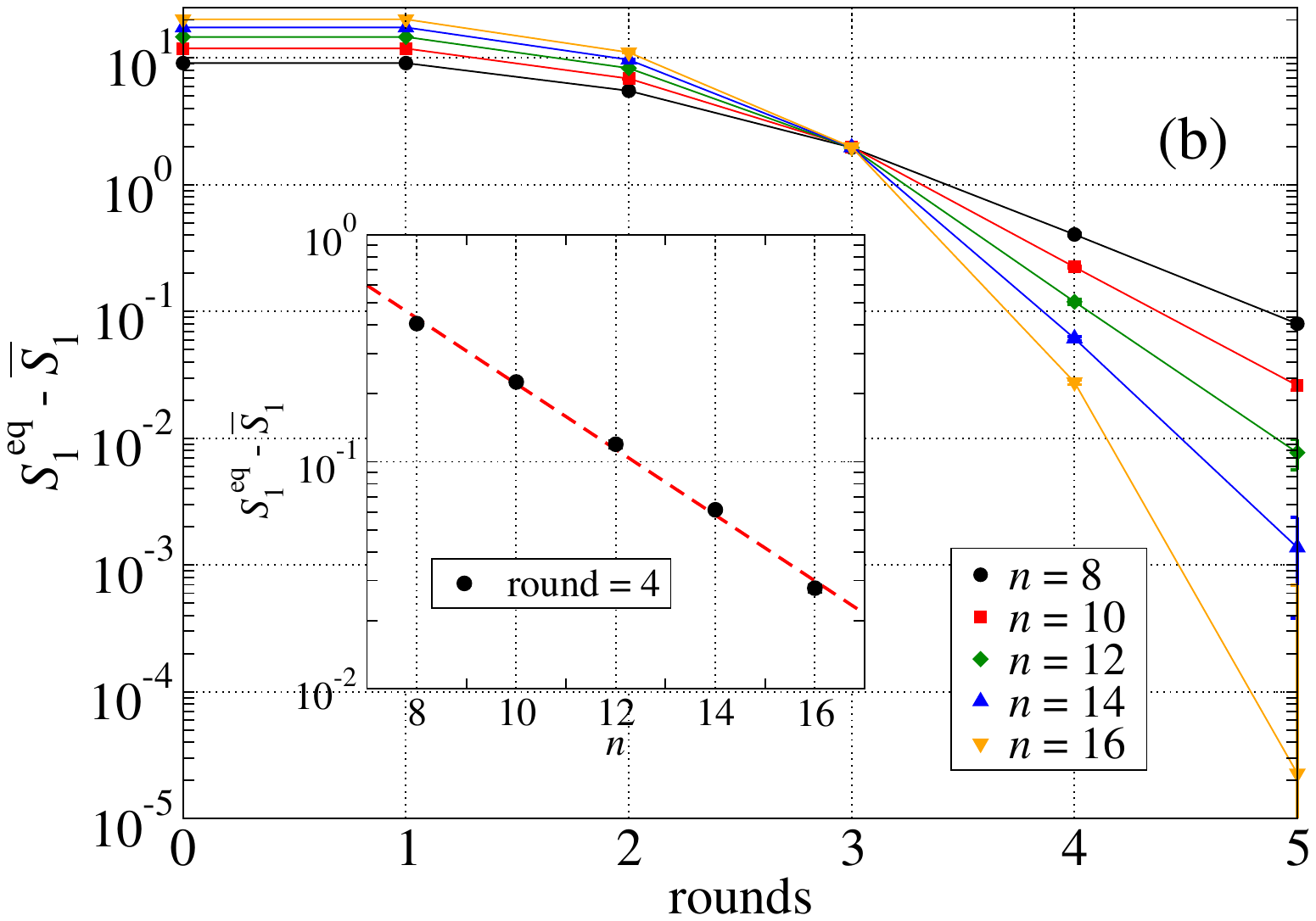}
\caption{(a) Average entropy, $\overline{S}_1$, over different
  realizations of Feistel ciphers, as a function of the number of
  rounds, for different block sizes, $n$. The initial string contains
  a single $\hat{\sigma}^{\rm x}$ on the $L$ register
  ($\alpha^{\rm z} = 0$ and $\alpha^{\rm x} = 2^{n/2-1}$). The number
  of Feistel ciphers used in the averaging range from 500 ($n=8$) to
  100 ($n=16$). Solid lines on the left panel are guides to the
  eye. The solid lines on the right sidebar panel show the equilibrium
  entropy values. (b) The deviation of entropy in panel (a) from the
  equilibrium value $S_1^{\rm eq}$. Notice that the deviations
  increase (decrease) with $n$ below (above) 3 rounds, whereas the
  deviation is independent of $n$ after exactly 3 rounds. The inset in
  (b) shows the decrease in the deviation from the equilibrium entropy
  after 4 rounds as a function of the number of bits. The dashed line
  corresponds to the fitted function
  $S_1^{\rm eq}-\overline{S}_1 = 6.2(9)\, e^{-0.33(1)n}$, where
  numbers in parentheses indicate standard errors. Pseudorandom
  functions used to build an $n$-bit cipher, $f_k$, $k=1,\dots, r$,
  are obtained by $r\;n/2$ sequential calls to a pseudorandom bit
  function with equal probability for outputs 0 or 1.}
\label{fig:S1_Feistel}
\end{figure}

\end{widetext}

Below we apply our criterium for a secure cipher based on the
saturation of the entropies to the Luby-Rackoff cipher. We compute the
entropies defined in Eq.~\eqref{eq:entropies-def} using the string
amplitudes $A_{\beta\alpha}$ in Eq.~\eqref{eq:explicit_A_from_tr},
calculated by using the permutations $P, P^{-1}$ of the Feistel cipher
for an initial string $\alpha$ that corresponds to a $\sx$ operator
(or NOT gate) on a bit on the $L$ register of the cipher. As shown in
Fig.~\ref{fig:S1_Feistel}(a), we find that the entropy $S_1$ marches
to its asymptotic equilibrium value $S_1^{\rm eq}$ [see
  Eq.~\eqref{eq:equil-S}] as the number of rounds
increases. Figure~\ref{fig:S1_Feistel}(b) displays the average (over
different realizations of ciphers) residual entropy $S_1^{\rm eq} -
{\overline S}_1$, and shows that the extensive contribution is
saturated for 3 or more rounds, as reflected in the crossing of the
curves for different block sizes at exactly 3 rounds. However, the
saturation of the remaining order 1 contribution to the asymptotic
equilibrium value, $\Delta S^{\rm eq}_1 = 1.9618961$, computed in the
previous subsection requires at least 4 rounds. The same saturation
behavior is illustrated in Table~\ref{table:1} for other entropies
($q\ne 1$). It is important to note that at 4 (or more) rounds $\Delta
S_1^{\rm eq} - \overline{\Delta S}_1$ vanishes exponentially in the
number of bits, $n$, as seen in the inset to
Fig.~\ref{fig:S1_Feistel}(b).

The analysis of the entropy saturation gives an intuitive
interpretation of the Luby and Rackoff results. Without knowledge of
any specific attack, the order 1 residual entropy of 3-round ciphers
underscores the vulnerability to an attack that can uncover of the
order of 1 bit of information. At the same time, the exponential decay
(with $n$) of the residual entropy of 4-round ciphers reflects the
fact that such ciphers are secure against any polynomial attack. We
note that while calculating the string entropy is exponentially
costly, the finite size scaling analysis (common in statistical
mechanics) establishes the above results by considering only
computationally accessible sizes.

\begin{widetext}
  \begin{center}
\begin{table}[h]
  \centering
  \begin{tabular}{c c c c c c c c c c c c c c}
    \hline
                 & \vline & round 1 & \vline & round 2 & \vline & round 3 & \vline & round 4 & \vline & round 5 & \vline & rand. perm. (numerical) \\
    \hline
    \hline
    $S_1^{\rm eq}-\overline{S}_1$ & \vline & 22.1807 & \vline & 11.08(1) & \vline & 1.98(1)  & \vline & 0.028(1)  & \vline & 0.00002(67)  & \vline & 0.00016(49) \\
    \hline
    $S_2^{\rm eq}-\overline{S}_2$ & \vline & 22.1807 & \vline & 11.07(1) & \vline & 2.7(1)  & \vline & 0.053(4)  & \vline & 0.0005(8)  & \vline & 0.00036(64) \\
    \hline
    $S_3^{\rm eq}-\overline{S}_3$ & \vline & 22.1807 & \vline & 11.05(2) & \vline & 4.1(3) & \vline & 0.124(14) & \vline & 0.0013(13) & \vline & 0.00064(87) \\
    \hline
  \end{tabular}
  \caption{Average residual entropies computed after rounds of a
    16-bit Feistel cipher. Initial states and values of $\alpha^{\rm
      z}$ and $\alpha^{\rm x}$ are the same as those used in
    Fig. \ref{fig:S1_Feistel}. 100 circuits are used to compute the
    average. Numbers in parentheses indicate standard deviations. The
    right-most column shows the residual entropy values for 16-bit
    random permutations computed numerically (average over 100
    permutations). The values of $S^{\rm eq}_q=16\ln 4 -\Delta S^{\rm
      eq}_q$ for the $q$ shown are computed from
    Eq.~\eqref{eq:equil-S} with the corrections in
    Eq.~\eqref{eq:delta_S_q}, explicitly $\Delta{S^{\rm
        eq}_1}=1.9618961$, $\Delta{S^{\rm eq}_2} = \ln 10$, and
    $\Delta{S^{\rm eq}_3} = \ln 14$.}
  \label{table:1}
\end{table}
\end{center}
\end{widetext}


\subsection{Orthogonal and Unitary transformations}
\label{sec:quantum-ensambles} 

The probability of the string amplitudes $A_{\beta\alpha}$ are much
easier to compute for orthogonal and unitary transformations, which,
unlike permutations, are continuous.

For the orthogonal case, the amplitudes are real and the distribution
is Gaussian, with the width determined by the normalization condition
in Eq.~\eqref{eq:norm}. The symmetry condition in
Eq.~\eqref{eq:symmetry_parity} restricts the set of string labels
$\alpha$ which lead to non-zero amplitudes to those satisfying the
parity condition $n_y(\alpha)=n_y(\beta)\!\!\mod 2$. This condition
eliminates amplitudes on half of the string space, and thus the
Gaussian distribution for the remaining $d/2$ states has width
$\sqrt{2/d}\;$ (recall that $d=4^n$):
\begin{align}
  p_{O}(A_{\beta\alpha})
  &=
  \sqrt{\frac{d}{4\pi}}\;e^{-\frac{d}{4} A_{\beta\alpha}^2}
  \;.
\end{align}

For the unitary case, the symmetry condition in
Eq.~\eqref{eq:symmetry_parity} dictates that the amplitudes are real
for $n_y(\alpha)=n_y(\beta)$ even, and purely imaginary for
$n_y(\alpha)=n_y(\beta)$ odd. By contrast to the orthogonal case, the
amplitudes are non-vanishing in all $d$ states, and thus the width of
the Gaussian distribution is $\sqrt{1/d}\;$:
\begin{align}
  p_{U}(A_{\beta\alpha})
  &=
  \sqrt{\frac{d}{2\pi}}\;e^{-\frac{d}{2} |A_{\beta\alpha}|^2}
  \;.
\end{align}

These probability distributions lead to the following order-1
corrections to the entropies:
\begin{align}
  \Delta{S^{O}_q}
  =&
  \ln 4 
  +
  \frac{1}{q-1}\;
  \ln\frac{\Gamma\left(q+\frac{1}{2}\right)}
           {\Gamma\left(\frac{3}{2}\right)}
\end{align}
and
\begin{align}
  \Delta{S^{U}_q}
  =&
  \ln 2 
  +
  \frac{1}{q-1}\;
  \ln\frac{\Gamma\left(q+\frac{1}{2}\right)}
           {\Gamma\left(\frac{3}{2}\right)}
\;.
\end{align}

\subsection{Vanishing of OTOCs at equilibrium}
\label{sec:OTOC}

For all three symmetry classes, the OTOCs vanish in the equilibrium
state. For independently and identically distributed amplitudes
$A_{\beta\alpha}$, the ensemble average of the expression in
Eq.~\eqref{eq:general-OTOC2} factorizes into products of clusters of
$2p$ $A$s. In each of these clusters the states $\gamma_i$ must be
equal, and the cluster contributes a factor of $\overline{A^{2p}}$. In
all of the three symmetry classes discussed above,
$\overline{A^{2p}}\sim d^{-p}$.

When there are two or more clusters, the number of independent strings
$\gamma$ to be summed over is $\sum_p n_p-1$, where $n_p$ is the
number of clusters of size $2p$, with the $-1$ reflecting the
reduction in the number of independent $\gamma$s by 1 due to the
Kronecker $\delta$ in Eq.~\eqref{eq:general-OTOC2}. It then follows
that the contribution to the OTOC from a given partition scales as
$d^{-1 + \sum_p n_p - \sum_p n_p p}$; and thus, OTOCs are dominated by
partitions with clusters of size 2 (or $p=1$), the contribution of
which vanishes as $d^{-1}$, the inverse of the size of the string
Hilbert space.  All other partitions contribute higher powers of
$d^{-1}$.  We note that the argument above did not take into account
the phase factors coming from the traces over the products of string
operators in Eq.~\eqref{eq:general-OTOC2}; these factors, which
introduce terms with alternating signs, can only further decrease the
OTOCs.

In the case of a single cluster of size $2p$, there is only one
$\gamma$ to be summed over, and the Kronecker $\delta$s in
Eq.~\eqref{eq:general-OTOC2} simply specify the $\alpha$s for which
the correlator is non-zero. When $p\ge 2$, the contribution to the
OTOC for one-cluster partitions, again not counting the phases (i.e.,
alternating signs) from the traces over the products of string
operators in Eq.~\eqref{eq:general-OTOC2}, vanishes as $d^{1-p}$,
i.e., at least as fast as $d^{-1}$. By contrast, for the case of a
single cluster with $p=1$ one must consider the random phases, the sum
over which, when weighted by the probabilities $A^2$, leads to an
additional factor of $\sqrt d$, and consequently, an OTOC decaying as
$d^{-\frac{1}{2}}$.

These arguments are common to the three symmetry classes for generic
string operators defining OTOCs. We note, however, that the case of
permutations contains an invariant subspace of the Hilbert space of
strings, namely the space of $z$-strings (these shall be further
discussed in Sec.~\ref{sec:dynamics}). An initial $z$-string,
containing a product of only Pauli $\hat \sigma_z$s and identities,
will evolve into a superposition of $z$-strings. The SAC OTOC of
Eq.~\eqref{eq:SAC} falls into this category in that the relevant string
amplitudes, $A_{\beta\gamma}(\tau)$, spread initial $z$-strings only
over a $\sqrt d=2^n$ subspace (instead of over the full
$4^n$-dimensional string space). Arguments exactly like those we made
above for the single cluster with $p=1$ still apply, but with the sum
over the $\gamma$'s restricted to the subspace with
$\gamma^{\rm x}=0$; this restriction implies that the OTOCs of this
type will decay as $d^{-\frac{1}{4}}$.

As will become clear in the next section, reaching an equilibrium
state of independently and identically distributed amplitudes
$A_{\beta\alpha}$ that result in the vanishing of arbitrary order
OTOCs, requires the use of a universal set of gates.

\section{String Dynamics}
\label{sec:dynamics}

For permutations, the dynamics of the string amplitudes is determined
by the specific implementation of $P$, for example, either through the
rounds of a Feistel cipher or through the application of a universal
set of classical gates. Here we focus on the evolution via a random
classical circuit decomposed in terms of 3-bit gates in the symmetric
group $S_8$, each of which leads to a transition amplitude in string
space of the form
\begin{align}
  \label{eq:hopping}
  t^{(g)}_{\alpha'\alpha}
  &=\bra{\alpha'}\;{\hat{\mathfrak g}}\;\ket{\alpha}
  \nonumber\\
  &=
  \frac{1}{2^n}\;
  \tr \left({\hat g}^\dagger \; {\widehat S}_{\alpha'}^{\;}\; {\hat g}\;
  {\widehat S}_{\alpha}^{\;\dagger}\;\right)
    \;,
\end{align}
where the string states $\alpha,\alpha'$ may differ only in a
substring of size 3. The transition amplitude
$t^{(g)}_{\alpha'\alpha}$ implemented by a single gate $g$ satisfies
the same normalization conditions as the $A_{\beta\alpha}$, namely:
$\sum_\alpha |t^{(g)}_{\alpha'\alpha}|^2=\sum_{\alpha'}
|t^{(g)}_{\alpha'\alpha}|^2=1$.

These matrix elements control the spreading of the string wave function
in the $d=4^n$ dimensional space of strings, analogous to the
spreading taking place in random quantum circuits that deploy random
unitary or orthogonal 2-qubit gates that are drawn from elements of
$U(4)$ or $O(4)$, respectively. The essential feature that controls
quantum diffusion and delocalization in string space is the
connectivity of a given string state, $\alpha$, to multiple states,
$\alpha'$, a feature which is common to all three gate-set symmetry
classes mentioned above. Below we discuss these three classes of
circuits -- those using 3-bit gates in $S_8$ or 2-qubit gates in
$U(4)$ or $O(4)$ -- on equal footing.

To quantify spreading, we will consider powers of the transition
probabilities $|t^{(g)}_{\alpha'\alpha}|^2$,
\begin{align}
  \label{eq:T_ave}
  T^q_{\alpha'\alpha}
  =
  \frac{1}{|G|}\sum_{g\in G}
  |t^{(g)}_{\alpha'\alpha}|^{2q}
  \;,
\end{align}
 averaged over the group $G$ for the given gate set, and the sum over
 these quantities over all final states $\alpha'$,
\begin{align}
  V^q_{\alpha}
  =
  \sum_{\alpha'} T^q_{\alpha'\alpha}
  \;,
\end{align}
which are measures of the connectivity of the state $\alpha$, averaged
over gates. Notice that $V^{q=1}_{\alpha} = 1$ follows from the
normalization condition $\sum_{\alpha'}|t^{(g)}_{\alpha'\alpha}|^2=1$,
and thus in diagnosing string spreading we must consider
$V^{q}_{\alpha}$ for $q\ne 1$. Since for a fixed initial state
$\alpha$ the $|t^{(g)}_{\alpha'\alpha}|^2$ are probabilities, it is
also useful to introduce an associated entropy
\begin{align}
  s_\alpha
  =
  -\frac{d}{dq} V_\alpha^q\Big|_{q\to 1}
  =
  -\sum_{\alpha'}
  \frac{1}{|G|}\sum_{g\in G}
  |t^{(g)}_{\alpha'\alpha}|^{2}\;\ln |t^{(g)}_{\alpha'\alpha}|^{2}
  \;.
\end{align}

As the simplest example, consider 2-bit gates in $S_4$ or 2-qubit
Clifford gates~\cite{Gottesman,Aaronson2004}, in which case
$|t^{(g)}_{\alpha'\alpha}|^2$ is a permutation matrix, i.e., it
connects single initial and final states, resulting in $V^{q}_{\alpha}
=1$ for all $q$s, and the vanishing of the entropy $s_\alpha$. By
contrast, for 3-bit gates in $S_8$ and 2-qubit unitary and orthogonal
gates, $V^{q}_{\alpha}\ne 1$ for $q\ne 1$, and $s_\alpha>0$. Table
\ref{tab:s-andVresult} displays the values of $V^{q\to 0}_{\alpha}$,
which measures the average connectivity of a state $\alpha$, and the
entropy $s_\alpha$. These results follow from the detailed structure
of the matrix $T^q_{\alpha'\alpha}$ for 3-bit gates in $S_8$, and
2-qubit gates in $U(4)$ and $O(4)$, shown in
Fig.~\ref{fig:blocks}. For all three sets of gates and in all
non-trivial subspaces, $V^{q\to 0}_{\alpha} > 1$ and $s_\alpha >
0$. It is then clear that string proliferation builds up exponentially
through the application of a sufficiently large number of consecutive
gates. It is this proliferation that evolves the entropies to their
saturation values and leads to vanishing OTOCs in the asymptotic
equilibrium state.

\begin{widetext}
\begin{center}
\begin{table}[h]
\centering
\begin{tabular}{c ccccc c cccc c ccc}
& \vline &\multicolumn{4}{c}{3-bit gates in $S_8$}
& \vline
& \multicolumn{3}{c}{2-qubit gates in $O(4)$}
& \vline
& \multicolumn{2}{c}{2-qubit gates in $U(4)$}
\\ [0.5ex]  
\hline\hline
Measure& \vline &\multicolumn{4}{c}{Sector}
& \vline
& \multicolumn{3}{c}{Sector}
& \vline
& \multicolumn{2}{c}{Sector}
\\ 
& \vline &
\;identity &\; z-strings &\; odd parity &\; even parity
& \vline &
\;identity &\; odd parity &\; even parity
& \vline
&\;identity &\; other strings
\\\hline
$V^{q\to 0}_\alpha$ & \vline &
1 & 17/5 &  51/5 & 103/10
& \vline
& 1 & 3 & 9
& \vline
& 1 & 15
\\ [0.5ex]
$s_\alpha$
& \vline &
0 & 1.11 & 2.03 & 2.08
& \vline
& 0 & 0.67 & 1.34
& \vline
& 0 & 1.90
\\ [1.5ex]
\hline
\end{tabular}
\caption{Connectivity and entropy measures for 3-bit permutation gates from $S_8$; and 2-bit gates in $O(4)$ and $U(4)$.}
\label{tab:s-andVresult}
\end{table}

\end{center}

\begin{figure}[h]
\includegraphics[width=0.9\textwidth]{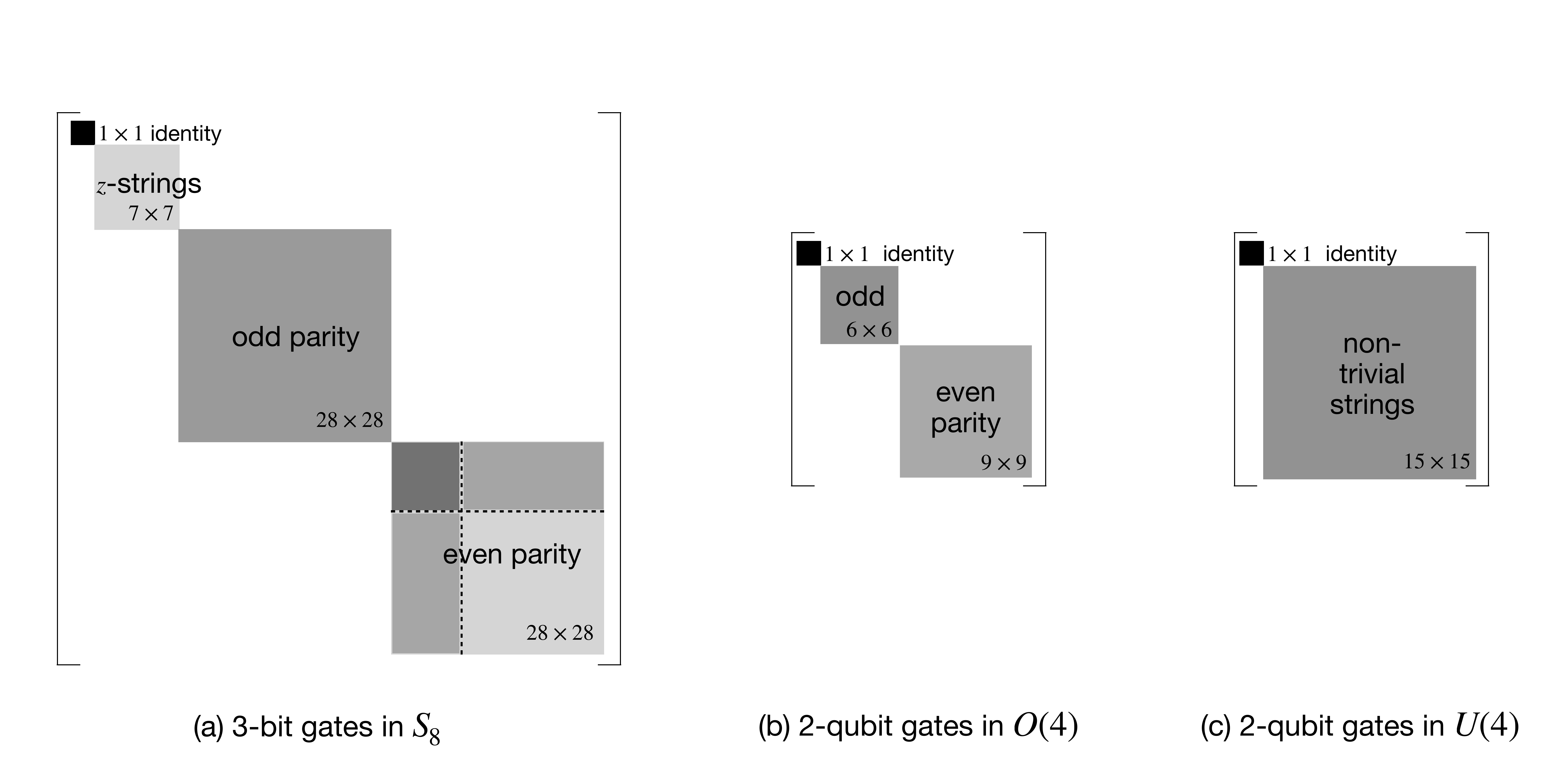}
\caption{
Block form for the matrices $T^q_{\alpha'\alpha}$ in
Eq.~\eqref{eq:T_ave} for the three symmetry classes. (a) For 3-bit
permutation gates, the $4^3\times 4^3$ matrix $T^q$ breaks up into 4
blocks of sizes $1\times 1$, $7\times 7$, $28\times 28$, and $28\times
28$, reflecting the symmetries of the string space transition
amplitudes. The $1\times 1$ block represents the transition between
identity strings -- tensor products of Pauli identity operators --
that are invariant under evolution. The $7\times 7$ block connects
only $z$-string states, i.e., those with $\alpha_x=0$, which form an
invariant subspace.
The remaining two $28\times 28$ blocks correspond to transitions
within subspaces with even or odd number $n_y(\alpha)$ of $\sy$ Pauli
matrices or, equivalently, the number of overlapping $\sx$ and $\sz$
Pauli matrices on the substring. Recall from the discussion in
connection with Eq.~\eqref{eq:symmetry_parity} that the parity of
$n_y(\alpha)$ is a conserved quantity.
The matrix elements of $T^q$ within each of the $1\times 1$, $7\times
7$, and odd parity $28\times 28$ blocks are equal. The even parity
$28\times 28$ block involves four sub-blocks: a diagonal $7\times 7$
block describing transitions between $x$-strings ($\alpha_z=0$);
another diagonal $21\times 21$ block of transitions between the
remaining odd parity state; and two off-diagonal, $7 \times 21$ and
$21\times 7$ rectangular blocks connecting these two subspaces. Within
each of these sub-blocks, the matrix elements are the same.
(b) For 2-qubit gates in $O(4)$, the $4^2\times 4^2$ matrix $T^q$
breaks up into the trivial $1\times 1$ block (connecting the identity
strings), a $6\times 6$ odd subsector, and a $9\times 9$ even
subsector. The matrix elements are all equal within each block. (c)
For the 2-qubit gates in $U(4)$, $T^q$ breaks into the identity block
and a $15\times 15$ block, with all equal matrix elements, connecting
all other states.
}
  \label{fig:blocks}
\end{figure}
\end{widetext}

A subtlety of the evolution governed by the application of gates drawn
uniformly from the universal set $S_8$ is the non-zero probability $p$
(independent of $n$) that a string of weight 1 does not increase even
when touched by a (dual) gate. [The same situation occurs for gates in
$O(4)$ and $U(4)$.] This ``stay-probability'' arises from the general
structure of the $T^q_{\alpha'\alpha}$, and implies that after $\ell$
layers this probability decreases to $\sim p^\ell$, corresponding to a
polynomial decay in $n$ for $\ell \sim\ln n$. Below we will use the
freedom afforded us in the context of ciphers to design circuits that
deploy different subsets of gates in $S_8$ in separate stages. These
multi-stage structure of circuits (of depth $\ell\sim \ln n$), which
eliminate the tails due to low weight strings, rely on separating two
distinct processes controlling the evolution of string wave function
amplitudes: (a) the extension of the weight of the string, which we
refer to as inflation; and (b) the proliferation of the number string
states with non-zero amplitudes. The latter, but not the former,
requires the action of nonlinear gates. Below we discuss these
distinct processes, {\it inflation} and {\it proliferation}, which
will be central to the structure of the optimal cipher proposed in
this paper.

\subsection{Inflation}
\label{sec:inflation}

The inflationary process is achieved through the action of a set of
special gates that eliminate the stay-probability for weight 1
substrings. We handpick these special gates as follows. First, we
compute the entries $t^{(g)}_{\alpha'\alpha}$ of the $64\times 64$
transition in Eq.~\eqref{eq:hopping} for each of the $8!$ gates in
$S_8$. We then filter out all gates for which the $9\times 9$
submatrix connecting the 9-dimensional subspace of weight 1 strings is
non-zero, thus retaining only gates for which weight 1 substrings grow
into weight 2 or 3 substrings. We identify 144 such ``inflationary''
gates, which are linear. These gates are listed explicitly in
Appendix~\ref{sec:app:inflationary}, where we also represent them in
terms of equivalent circuits of CNOTs.

To gain further insight into the effect of the 144 inflationary gates,
we note that, in bit space, this set of gates has the following
property: for any input state, flipping one input bit flips at least 2
output bits. It follows that, after $\ell$ layers, flipping one of the
$n$ input bits of a circuit leads to a cascade of $2^\ell$ flipped
output bits, and thus $\ell=\log_2 (n/2)$ layers suffice to change
half of the output bits. The cipher proposed in Sec.~\ref{multi-stage}
employs a tree-structured circuit (see below) for which such a cascade
occurs, thus eliminating tails in the stay probability of low-weight
strings.

\subsection{Proliferation}
\label{sec:proliferation}

As already discussed in connection with Table~\ref{tab:s-andVresult},
the increase in entropy is connected with the proliferation of strings
induced through the action of nonlinear gates in $S_8$.  In order to
speed up entropy production, instead of deploying the full set of $8!$
gates in $S_8$, we use the subset of nonlinear gates that maximizes
string proliferation at each step. This set is identified from the
entries $t^{(g)}_{\alpha'\alpha}$ for every gate $g$ by selecting
those for which the number of non-zero elements in each line or column
is maximal. We find 10752 ``super-nonlinear'' gates satisfying this
criterium. For all these gates the squares of the non-zero elements
take the same values within each of the blocks shown in
Fig.~\ref{fig:blocks}. In each line or column there are: 4 equal
values of 1/4 in the $7\times 7$ $z$-block; and 16 equal values of
1/16 in both the $28\times 28$ even and odd blocks. This structure
implies that each of these gates leads to string proliferation with
each and every application. From the scattering matrix elements one
also determines the values of $V^{q\to 0}_{\alpha}$ that measure the
average connectivity of a state $\alpha$ and the entropy $s_\alpha$,
shown in Table~\ref{tab:s-andVresult-super-nonlinear} (notice that
these values are -- through the design choice above -- larger than the
averages for the gates in $S_8$ in Table~\ref{tab:s-andVresult}.) We
established that this super-nonlinear subset of gates constitute a
universal set for classical computation by determining that the order
of the group generated by particular pairs of gates in this subset is
8!~\footnote{For example, the two permutations $0\, 1\, 2\, 4\, 3\,
6\, 7\, 5$ and $7\, 3\, 0\, 5\, 1\, 2\, 4\, 6$ suffice to generate
$S_8$.}

\begin{center}
\begin{table}[h]
\centering
\begin{tabular}{c ccccc c}
& \vline &\multicolumn{4}{c}{Super-nonlinear subset of gates in $S_8$}
& \vline
\\
\hline\hline
Measure& \vline &\multicolumn{4}{c}{Sector}
& \vline
\\ 
& \vline &
\;identity &\; z-strings &\; odd parity &\; even parity
&\vline
\\\hline
$V^{q\to 0}_\alpha$ & \vline &
1 & 4 &  16 & 16
& \vline
\\ [0.5ex]
$s_\alpha$
& \vline &
0 & $2\ln 2$ & $4\ln 2$ & $4\ln 2$
& \vline
\\ [0.5ex]
\hline
\end{tabular}
\caption{Connectivity and entropy measures for super-nonlinear gates,
  a subset (with cardinality 10752) of the set of 8! gates in $S_8$.}
\label{tab:s-andVresult-super-nonlinear}
\end{table}

\end{center}


Particularly relevant to our discussion below is the action of
super-nonlinear gates on typical macroscopic weight strings generated
via inflation. In the evolution of such strings with large cross
section, the action of every super-nonlinear gate leads to a
superposition of strings and thus, every layer of $n/3$
super-nonlinear gates increases the number strings exponentially in
$n$, resulting in a macroscopic increase in the string entropy. We
note, however, that the exponential proliferation of strings already
induced by the action of one layer of nonlinear gates is, by itself,
not sufficient to ensure the equilibration of the system. As will be
explained in Sec.~\ref{multi-stage} in the context of multi-stage
tree-structured circuits, in order to reach the asymptotic probability
distribution of string amplitudes discussed in
Sec.~\ref{sec:equilibrium}, one requires a minimum number of layers of
both inflationary and super-nonlinear gates. This number is determined
by the accuracy to which one requires the vanishing of the OTOCs or
the saturation of the entropies, criteria which determine the security
of the cipher.

\section{Multi-stage cipher}
\label{multi-stage}

\begin{figure}[h]
\centering
\includegraphics[width=0.45\textwidth]{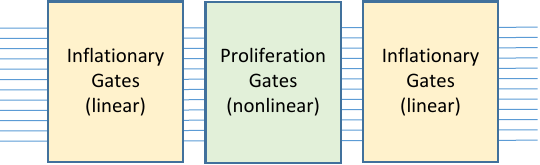}
\caption{A three-stage cipher, with two stages of inflationary gates
  flanking a stage of (super-)nonlinear gates.}
\label{fig:3-stage}
\end{figure}

Below we will exploit the features of the special inflationary and
super-nonlinear gates described above to introduce a three-stage
fast-scrambling cipher, the structure of which is illustrated in
Fig.~\ref{fig:3-stage}.

More precisely, we break up the operator $\hat P$ into three stages,
\begin{align}
  \label{eq:P_3-stage}
  \widehat P = \widehat L_r\;\widehat N\;\widehat L_l
\end{align}
where the $\widehat L_{l,r}$ are two different circuits composed of
inflationary gates, placed to the left and right of a circuit
$\widehat N$ of super-nonlinear gates, as depicted in
Fig.~\ref{fig:3-stage}. (The connectivity of these circuits are
tree-structured as discussed below.) The general OTOC of
Eq.~\eqref{eq:general-OTOC} can be re-expressed as
\begin{widetext}
  \begin{align}
    \label{eq:re-written-OTOC}
    C^{\,\alpha_1,\beta_1\dots\alpha_k,\beta_k}_{\rm CPCA} &= \tr\left[
    \rho_\infty\;
    \S^N_{l(\alpha_1)}(0)\;\S^N_{r(\beta_1)}(\tau)\;
    \S^N_{l(\alpha_2)}(0)\;\S^N_{r(\beta_2)}(\tau)\;
    \dots\;
    \S^N_{l(\alpha_k)}(0)\;\S^N_{r(\beta_k)}(\tau)
    \right] \;,
\end{align}
\end{widetext}
where
\begin{align}
  \label{eq:N-evolution}
  {\widehat S}^N_{r(\beta)} (\tau)\equiv
  \widehat N^\dagger \;{\widehat S}_{r(\beta)}\;\widehat N
  \quad
  \text{and}
  \quad
  {\widehat S}^N_{l(\alpha)}(0)\equiv
  {\widehat S}_{l(\alpha)}
  \;,
\end{align}
with
\begin{align}
  {\widehat S}_{r(\beta)}\equiv
  \widehat L_r^\dagger
  \, {\widehat S}_\beta\,
  \widehat L_r^{\;}
   \quad
  \text{and}
  \quad
  {\widehat S}_{l(\alpha)}\equiv
  \widehat L_l^{\;}
  \, {\widehat S}_\alpha\,
  \widehat L_l^\dagger
  \;.
\end{align}
The inflationary (linear) circuits $\widehat L_{l,r}$ map single
strings labeled by $\alpha$ and $\beta$ onto single strings labeled by
$l(\alpha)$ and $r(\beta)$, respectively. (We abuse notation and use
$l,r$ defining the specific inflationary circuits to also label the
respective functions that map initial into final strings.)

The rewriting of the OTOC in Eq.~\eqref{eq:re-written-OTOC} - in terms
of products of large weight strings resulting from the inflation of
the original initial and final state strings via layers of
inflationary linear gates, which evolve through conjugation by the
operator $\widehat N$, defined in Eq.~\eqref{eq:N-evolution} -
provides an intuitive picture for the vanishing of OTOCs. The
evolution of an inflated string from the output end with each layer of
$\widehat N$ yields an exponential number of strings, ${\cal N}_s$,
that spread over the $d$-dimensional Hilbert space of
strings. Following the arguments in Sec.~\ref{sec:OTOC}, OTOCs are
suppressed, at the very least, as the quartic root of ${\cal N}_s$,
thus yielding OTOCs that vanish exponentially in $n$. This intuitive
picture is borne out by an explicit calculation presented in
Sec.~\ref{sec:sac-recursion} for the simplest SAC OTOC, using the
tree-structured circuits of
Sec.~\ref{tree-structured-cipher}. Moreover, this explicit calculation
is consistent with the final conclusion of the paper (see below),
namely, that ${\cal O}(\ln n)$ layers of gates, for each of the
inflation and proliferation stages, are sufficient to suppress all
OTOCs exponentially in $n$.

\subsection{Tree-structured circuits}
\label{tree-structured-cipher}

Below we describe in detail circuits mentioned in the introduction, in
which triplets of bits acted upon by 3-bit gates are arranged in a
hierarchical (tree) structure. We consider the case when $n$ is a
power of 3, $n=3^q$. We proceed by forming groups of triplets of
indices for each layer, selected as follows:
\begin{align}
  \ell = 1: & \quad(0,1,2)\;(3,4,5)\;(6,7,8) \dots \nonumber\\
  \ell = 2: & \quad(0,3,6)\;(1,4,7)\;(2,5,8) \dots \nonumber\\
  \ell = 3: & \quad(0,9,18)\;(1,10,19)\;(2,11,20) \dots \nonumber\\
  \ell = 4: & \quad(0,27,54)\;(1,28,55)\;(2,29,56) \dots \nonumber\\
  \dots \quad&
\end{align}
More precisely, each of the $n/3=3^{q-1}$ triplets in layer $\ell$ are
indexed by $(i,j,k)$, which we write in base 3 as
\begin{align}
  i=&z_0 + 3\;z_1+3^2\;z_2+\dots+3^{\ell-1}\times\underline{0}+\dots\;3^{q-1}\;z_{q-1}\nonumber\\
  j=&z_0 + 3\;z_1+3^2\;z_2+\dots+3^{\ell-1}\times\underline{1}+\dots\;3^{q-1}\;z_{q-1}\nonumber\\
  k=&z_0 + 3\;z_1+3^2\;z_2+\dots+3^{\ell-1}\times\underline{2}+\dots\;3^{q-1}\;z_{q-1}
      \;,
      \label{eq:trit-triples}
\end{align}
where $z_a=0,1,2$, for $a=0,\dots, q-1$. Notice that at layer $\ell$
the members of the triplets, $(i,j,k)$, are numbers that only differ
in the $(\ell-1)$-th trit, while the other $q-1$ trits
$z_a, a\ne \ell-1$, enumerate the $3^{q-1}=n/3$ triplets. (If more
than $q$ layers are needed, we recycle in layer $\ell > q$ the
triplets of layer $\ell\!\!\mod q$.)

Once the triplets of indices, $(i,j,k)$, are selected for each layer,
we map them onto groups of three bits indexed by,
$\left(\pi(i),\pi(j),\pi(k)\right)$, via a (randomly chosen)
permutation $\pi$ of the $n$ bitlines. We note that for these
tree-structure ciphers, the key consists of the data needed to specify
the circuit, namely: (i) the permutation $\pi$; and (ii) the list of
gates in $S_8$ chosen to act on each of the triplets, for all
layers. This key uniquely defines the circuit, and its inverse.

\subsection{String inflation with a tree-structured circuit}
\label{sec:inflation-with-tree}

The tree-structured gate arrangement makes it rather simple to analyze
the string inflation process. Let us consider the conjugation of the
string operator ${\widehat S}_\alpha$ with the inflationary circuit
$\widehat L_l^{\;}$,
${\widehat S}_{l(\alpha)}\equiv \widehat L_l^{\;} \, {\widehat
  S}_\alpha\, \widehat L_l^\dagger$. The conjugated string operator,
${\widehat S}_{l(\alpha)}$, can be written with the aid of
Eq.~\eqref{eq:string-alpha} as
\begin{align}
  {\widehat S}_{l(\alpha)}
  &
    =
    \widehat L_l^{\;} \left(
    \prod_{j\in\alpha^{\rm x}}\;\sx_j
    \;
    \prod_{k\in\alpha^{\rm z}}\;\sz_k
    \right) \widehat L_l^\dagger
    \nonumber\\
  &=
    \prod_{j\in\alpha^{\rm x}}\;
    \left(\widehat L_l^{\;}\;\sx_j\;\widehat L_l^\dagger\right)
    \;
    \prod_{k\in\alpha^{\rm z}}\;
    \left(\widehat L_l^{\;}\;\sz_k\;\widehat L_l^\dagger\right)
    \;,
    \label{eq:L-conjugation-string}
\end{align}
whereby the inflation of the whole string is expressed in terms of the
inflation of individual Pauli operators, on which we concentrate
hereafter.

Each of the inflationary gates in a layer of $\widehat L_l^{\;}$
overlaps with a substring of length 3, and changes the weights of the
substring upon conjugation, according to rates extracted from the
transition matrices $t^{(g)}_{\alpha'\alpha}$ for gates $g$ in the
inflationary set. A weight 0 substring remains unchanged under
conjugation. The conjugation of a single Pauli operator with the
gate-operator $\hat g$ generates a product of either 2 or 3 Pauli
operators, i.e., the initial weight 1 string expands to weight 2 or
3. (The expansion to weight 2 or 3 depends on the relative position of
the initial Pauli operator with respect to the three bitlines on which
the inflationary gate $g$ acts; in 2 out of 3 cases it expands to
weight 2, and in 1 out of 3 cases to weight 3.) There are also matrix
elements for transitions of weight $2\to 1$ (2 out of 3 cases) and
$2\to 2$ (1 out of 3 cases), as well as $3\to 1$.

Using these transition rates, averaged over the gates in the
inflationary set (which homogenizes effects of relative position
between the Pauli operators on the string and the three bitlines), we
carry out an analysis of the string growth within a mean-field
treatment on the tree-structured circuit. Let us define the string
density $\rho$ of non-trivial Pauli operators as the weight $w$ of a
string divided by $n$. As a result of conjugation by a 3-bit
inflationary gate, one obtains a recursion relation relating the
string densities $\rho(\ell+1)$ and $\rho(\ell)$ in consecutive
layers:
\begin{align}
  \label{eq:MF-3bit}
  \rho(\ell+1)
  =&\;\;\;\frac{1}{3}\left\{
     [1-\rho(\ell)]^3 \times 0
     \right.
  \nonumber\\
  &\quad+
    \rho(\ell)\;[1-\rho(\ell)]^2 \left[2 \times 2 + 3\times 1\right]
  \nonumber\\
  &\quad+
  [\rho(\ell)]^2\;[1-\rho(\ell)] \left[1\times 2 + 2\times 1\right]
  \nonumber\\
   &\quad\left.
     +[\rho(\ell)]^3 \times 1
     \right\}
  \nonumber\\
  =&\;\;\;
     \frac{7}{3}\,\rho(\ell)
     -
     \frac{10}{3}\,[\rho(\ell)]^2
     +
     \frac{4}{3}\, [\rho(\ell)]^3
  \;.
\end{align}
In deriving this recursion we assume that the contributions to the
weight of the substring due to each of the three bitlines acted upon
by a gate are uncorrelated and uniform -- for example, the probability
that the substring has weight 3 is $[\rho(\ell)]^3$.

\begin{figure}[htp]
    \centering
      \includegraphics[angle=0,scale=.34]{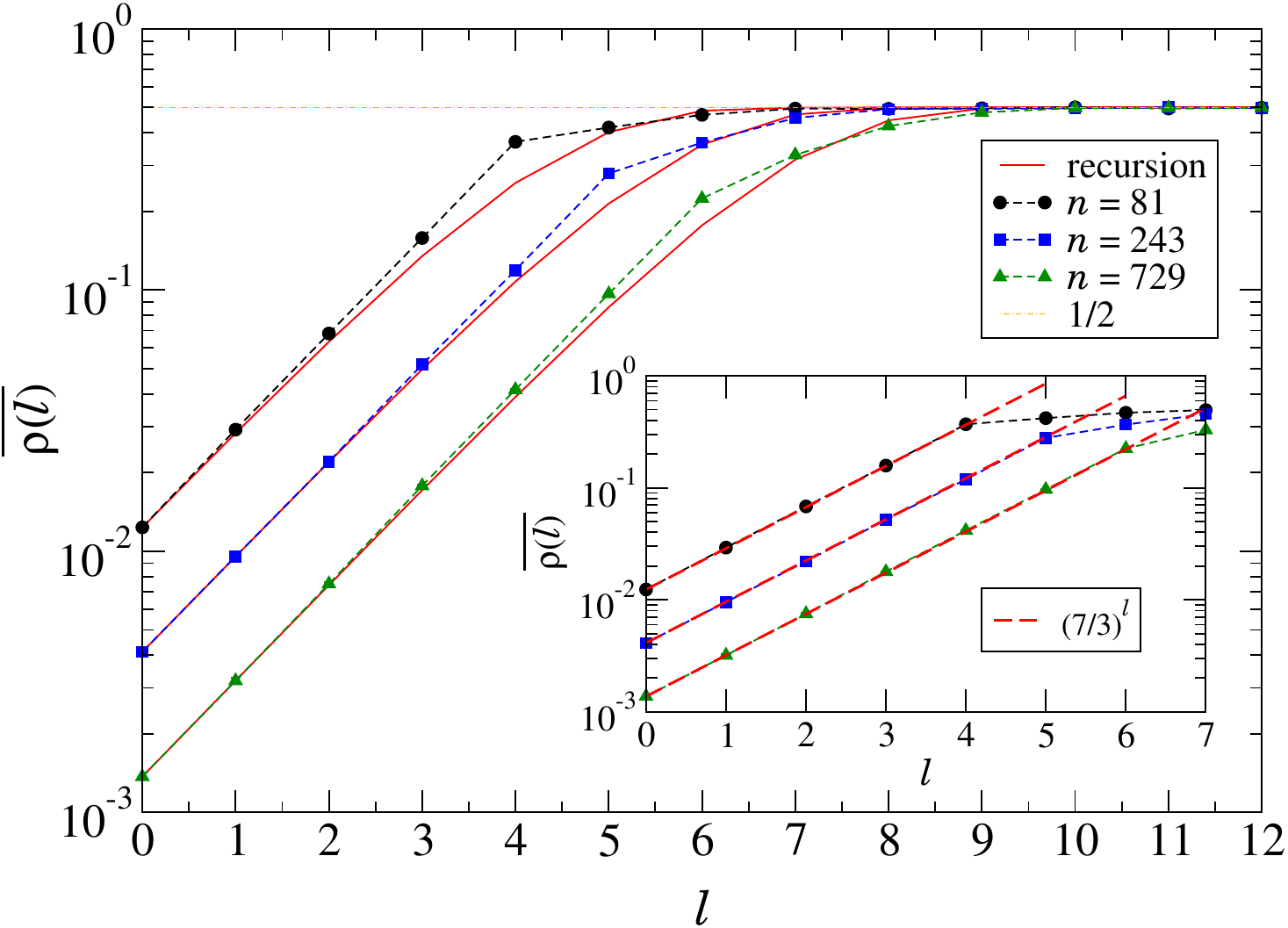}
      \caption{Average density of NOT gates ($\sx$) as a function of
        the number, $\ell$, of conjugation layers of inflationary
        gates for $n=81$, 243, and 729 (dashed lines are guides to the
        eye). The red lines indicate the recursion prediction,
        Eq. (\ref{eq:MF-3bit}), with the initial condition
        $\rho(0) = 1/n$. The dashed-dotted line indicates the
        saturation value $1/2$. Averages are taken over 512
        independent realizations of tree-structured circuits assembled
        by randomly and uniformly drawing from the 144 inflationary
        gates, and permuting the $n$ bitlines, as described in
        Sec.~\ref{tree-structured-cipher}. Statistical error bars are
        too small to be discerned. The inset shows the agreement with
        the initial growth of the density scaling as $(7/3)^\ell$
        (long dashed lines). Note also that the saturation at
        $\rho = 1/2$ after $\ell \sim \log_2 n$ layers predicted by
        the analytical mean field calculation is consistent with the
        qualitative argument given in the last paragraph of
        Sec.~\ref{sec:inflation}. As expected, the deviations (at
        intermediate densities) of the numerical simulations from the
        mean-field prediction in Eq. (\ref{eq:MF-3bit}) decrease as
        $n$ increases.  }
      \label{fig:NOT_N_g}
\end{figure}

The recursion Eq.~\eqref{eq:MF-3bit} has a fixed point at $\rho=1/2$,
which encodes that a single Pauli operator $\sx, \sy$ or $\sz$ on a
given bitline would eventually evolve into a string of $\sx$s, $\sy$s,
or $\sz$s on half the bitlines. The approach to this asymptotic
density is better captured by defining the quantity $m=2\rho-1$, and
studying how fast it approaches 0. The recursion relation for $m$
reads
\begin{align}
  \label{eq:s-recursion-inflation-section}
  m(\ell+1)
  = \frac{2}{3}\,[m(\ell)]^2 + \frac{1}{3}\,[m(\ell)]^3
  \;.
\end{align}
We highlight the absence of a term linear in $m(\ell)$ on the
right-hand side of Eq.~\eqref{eq:s-recursion-inflation-section}, which
makes the decay of $m$ towards zero a double exponential in $\ell$
that, for $\ell\sim {\cal O}{(\log n)}$, translates into an
exponential decay in $n$. Had we deployed the full set of gates in
$S_8$, the recursion relation would instead contain a linear in
$m(\ell)$ term, slowing down the approach to the asymptotic value of
the string density to exponential in $\ell$ and thus polynomial in $n$
for similarly sized circuits, an issue connected to the
stay-probabilities of string weights that we discussed above. (We
return to the discussion of the presence of such linear terms in
recursion relations derived for the SAC OTOC in
Sec.~\ref{sec:sac-recursion}, where we also discuss the case of
super-nonlinear gates.) Our analytical arguments lead to results in
agreement with those obtained from numerical simulations, as shown in
Fig.~\ref{fig:NOT_N_g}.

Henceforth, we will fix the number of layers $\ell_{L_l}$ and
$\ell_{L_r}$ of the inflationary stages $\widehat L_{l}$ and
$\widehat L_{r}$, respectively, of our cipher to be
$\ell_{L_l}=\ell_{L_r}=\log_2 n$, which suffices to reach the
saturation density $\rho=1/2$ and, more importantly, will also lead to
exponential decay (in $n$) of OTOCs, as shown below in
Sec.~\ref{sec:sac-recursion}.

\subsection{String proliferation with a tree-structured circuit}
\label{sec:proliferation-with-tree}

Here we exploit the tree-structured circuit to estimate the rate of
string-entropy production as a function of the number of layers of
super-nonlinear gates employed in $\widehat N$. The conjugation of the
macroscopic string $\widehat S_{l(\alpha)}$ generated by inflation is
carried out layer-by-layer, gate-by-gate of $\widehat N$.

We proceed by determining the entropy produced via conjugation with
one layer of super-nonlinear gates. Because of the macroscopic
cross-section of inflated strings, every one of the $n/3$
super-nonlinear gates of the layer will produce, on average, entropy
$\Delta s$. Averaging the entries for $s_\alpha$ in
Table~\ref{tab:s-andVresult-super-nonlinear} over the sectors yields
$\Delta s = 7/64\times 2\ln 2 + 28/64\times 4\ln 2 + 28/64\times 4\ln
2\approx 2.58$. In turn, the total string-entropy produced by
conjugation with the $n/3$ gates of one layer of $\widehat N$ is
$S^{N}(1)\sim\Delta s\times {\frac{n}{3}}$.

Because of the hierarchical (tree-structured) wiring of the circuit,
the entropy generated by conjugation with consecutive layers should be
additive, and therefore the string-entropy after $\ell$ layers scales
as $S^{N}(\ell)\sim S^{N}(1)\times \ell$, up to the point where
$S^{N}(\ell)$ curves to account for the saturation to its maximum
equilibrium value $S_1^{eq}=n\,\ln 4 -\Delta_1^{\rm eq}$ (see
Sec.~\ref{sec:random_permutations}). The linear scaling with $\ell$
allows us to conclude that producing the maximum extensive part of the
entropy, $n\ln 4$, only requires ${\cal O}(1)$ layers, namely
$\ell_s = 3\,\ln 4/\Delta s \approx 1.61$. It is the process of
saturating the {\it sub-extensive} correction to its precise value
$\Delta_1^{\rm eq}$ that requires ${\cal O}(\log n)$ layers!

The circuit depth for saturation of the sub-extensive correction to
the entropy can be determined from the following argument. Because of
the tree-structure of the circuit, after $q-1$ layers, one can
identify three independent subsystems that are not yet connected by
super-nonlinear gates. The three disjoint subset of bits comprising
these subsystems can be grouped according to the value of the most
significant trit indexing bitlines [see Eq.~\eqref{eq:trit-triples}].
(It is not until layer $\ell=q$ that these three separate subsystems
are connected by super-nonlinear gates.) The maximum entropy for the
system that can be reached with $q-1$ layers is the sum of the
saturation entropies of the three subsystems -- additive because of
separability -- namely,
$3\times[(n/3)\,\ln 4-\Delta_1^{\rm eq}]=n\,\ln 4-3\,\Delta_1^{\rm
  eq}$, a value $2\,\Delta_1^{\rm eq}$ below the equilibrium entropy
of the whole system. It then follows that, in order to eliminate this
entropy deficit and saturate the entropy, including its sub-extensive
universal correction of ${\cal O}(1)$, one needs at least the
additional $q$th layer of super-nonlinear gates to bridge the three
substructures and connect all $n=3^q$ bits of the system. We
conjecture that because of the extensive rate of entropy production
per layer, once this bottleneck is eliminated by connecting across all
degrees of freedom in the system, the full saturation of the entropy
to its maximum is achieved.

We remark that the saturation of the entropy with the application of
${\cal O} (\log n)$ layers of super-nonlinear gates is consistent with
the following observation. The degree of the Boolean polynomial
functions representing individual output bits of a circuit built out
of $\ell _N$ layers of nonlinear gates is at most $2^{\ell _N}$
because 3-bit gates output are at most quadratic functions. Therefore,
$\ell _N \sim {\cal O}(\log n)$ layers of nonlinear gates are required
in order to reach pseudo-random permutations, for which individual
outputs are polynomials of arbitrary order up to $n$. Generically,
such Booleans are unlearnable in polynomial time.

\subsection{SAC OTOC for the multi-stage cipher}
\label{sec:sac-recursion}

The tree structure enables a calculation of the SAC OTOC,
Eq.~\eqref{eq:SAC}, which shows explicitly that the cipher structured
as in Fig.~\ref{fig:3-stage}, with three stages each built out of
$\log_2 n$ layers, suppresses the OTOC exponentially in $n$. The case
of the SAC OTOC provides the simplest illustration of the mechanism by
which the three-stage cipher is secure against differential attacks.

We compute the square of the SAC OTOC,
$q^{ij}\equiv\left(C^{ij}_{\rm SAC}\right)^2$, as a function of the
number of applied layers of gates $\ell$ of the permutation associated
with the operator $\widehat P$ in Eq.~\eqref{eq:P_3-stage}. The
calculation, which is carried out in bit space, makes the mean-field
assumption (checked {\it a posteriori} in a numerical simulation) that
the system self-averages and $q^{ij}=q$, independent of $i$ and
$j$. We structure the calculation recursively, layer-by-layer,
relating $q(\ell+1)$ to $q(\ell)$. We note that the recursion relation
depends on the gate content (inflationary or super-nonlinear) of each
layer.

\begin{figure}[h]
\centering
\includegraphics[width=0.3\textwidth]{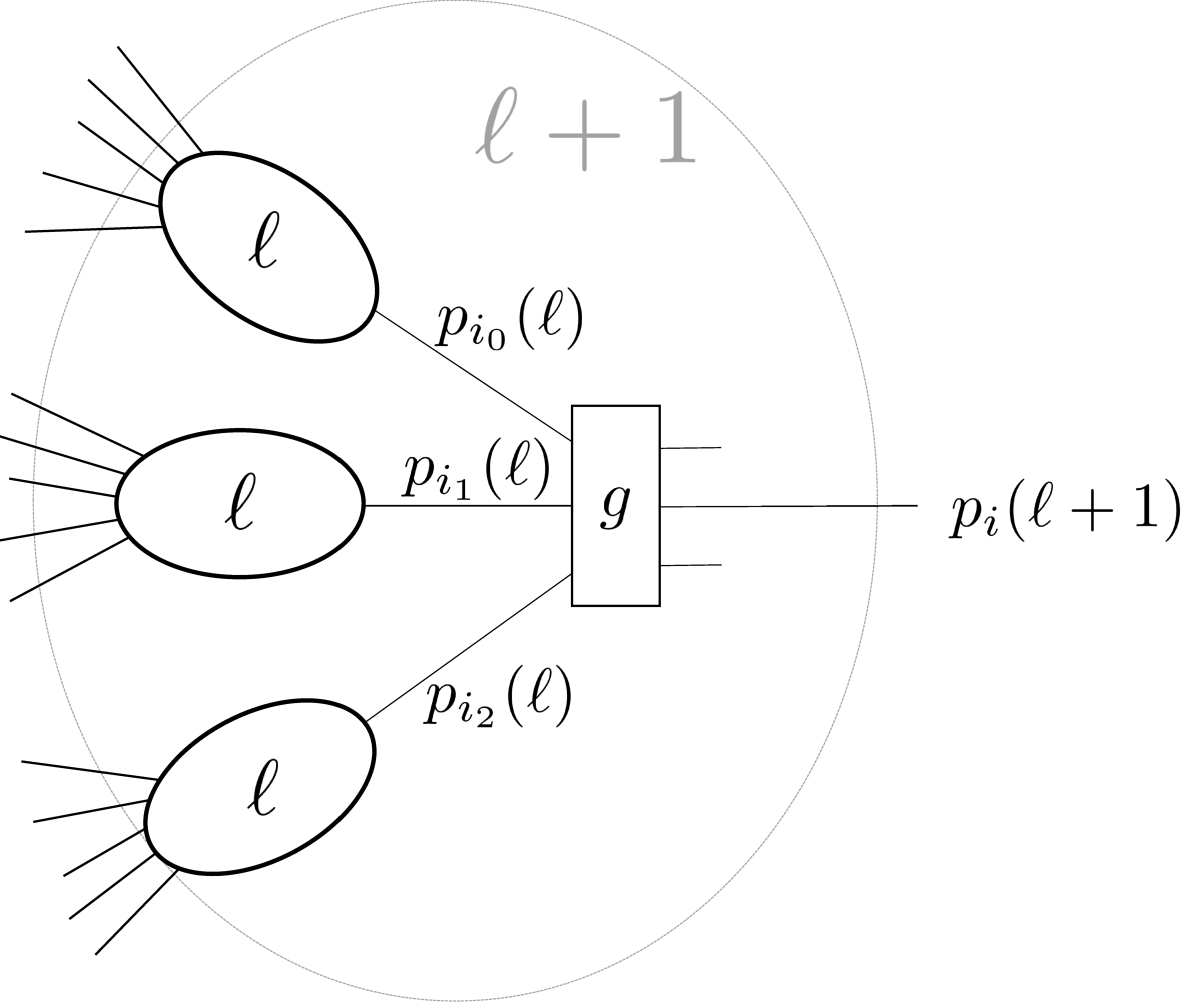}
\caption{The hierarchical structure of the circuit connectivity that
  illustrates the arguments used in the derivation of a recursion
  relation for the probability $p_i$ that a bit flips upon flipping a
  number of inputs.}
\label{fig:tree}
\end{figure}

We proceed by defining $p_i{(\ell)}$ to be the fraction of all
possible input states such that $x^{\rm out}_i$ remains put at
$x^{\rm out}_i$ [as opposed to flipping to $\overline{x^{\rm out}_i}$,
with probability $1-p_i{(\ell)}$] upon flipping a set of inputs of a
system with $\ell$ layers. In the hierarchical tree construction, a
given output bit $i$ (at level $\ell+1$) is obtained by taking three
outputs, at bitlines $i_0,i_1, i_2$ coming from separate branches of
the tree (at level $\ell$), as inputs to a 3-bit gate $g$ that affects
the output bit $i$ (see Fig.~\ref{fig:tree}). The specific action of
the gate $g$ determines the fraction of inputs for which the output
$i$ flips when $x\to x\oplus c$, with
$x\equiv x_{i_0} + 2\;x_{i_1} + 2^2\,x_{i_2}$ and
$c\equiv c_0+2\;c_1+2^2\,c_2$
encoding which ones of the three bits are flipped ($c_{0,1,2}=0$ for
an unflipped input or 1 for a flipped one). This fraction in expressed
as a coefficient
$C^{g_i}_{c_0c_1c_2}\equiv C^{g_i}_{c}=2\,f^{g_i}_{c}-1$, with
\begin{align}
  f^{g_i}_{c}
  =
  \frac{1}{2^3}\sum_{x=0}^7 (-1)^{{g_i}(x)\oplus {g_i}(x\oplus c)}
  \;.
\end{align}
\begin{widetext}
  The recursion for the flipping probabilities can then be expressed as
\begin{align}
  \label{eq:recursion_pi}
  p_i{(\ell+1)}
  &=
    h\left(p_{i_0}{(\ell)}, p_{i_1}{(\ell)}, p_{i_2}{(\ell)}; \{C^{g_i}_c\}\right)
  \\
  &=
  p_{i_0}{(\ell)}\,p_{i_1}{(\ell)}\,p_{i_2}{(\ell)}\;C^{g_i}_{000}
  +
  (1-p_{i_0}{(\ell)})\,p_{i_1}{(\ell)}\,p_{i_2}{(\ell)}\;C^{g_i}_{100}
  +
  \cdots
  +
    (1-p_{i_0}{(\ell)})\,(1-p_{i_1}{(\ell)})\,(1-p_{i_2}{(\ell)})\;C^{g_i}_{111}
    \;.
    \nonumber
\end{align}
Note that Eq.~\eqref{eq:recursion_pi} refers to a gate, $g$, which
is part of a specific, given realization of a circuit. We now proceed
to consider ensembles of circuits, and analyze the evolution of the
probability distribution, $P(p_i;\ell)$, of the $p_i$, as function of
$\ell$. The recursion relation for $P(p_i;\ell)$ is obtained by using
Eq.~\eqref{eq:recursion_pi}, and reads
\begin{align}
  P{(p_i;\ell+1)}
  &=\sum_{g\in S_8}
    \int dp_{i_0}\,dp_{i_1}\,dp_{i_2}\;
    P(p_{i_0};\ell)\;P(p_{i_1};\ell)\;P(p_{i_2};\ell)\;
    {\cal P}_{\rm set}(g)\;
    \delta\left[p_i-
    h\left(p_{i_0}, p_{i_1}, p_{i_2}; \{C^{g_i}_c\}\right)
    \right]
    \;,
\end{align}
where the gates are drawn from a probability distribution
${\cal P}_{\rm set}(g)$, which depends on the gate set, and we assumed
that the distribution $P$ is independent of the bitline index $i$. The
initial condition is determined by the fraction $f$~\footnote{We note
  that the assumption of independence of the bitline index cannot be
  justified unless $f$ is intensive, which only occurs through the
  action of sufficient number of layers of inflationary gates, see
  Fig.~\ref{fig:NOT_N_g}.} of bits that are flipped on input:
\begin{align}
  \label{eq:initial-P}
  P{(p;\ell=0)}
  =
  f\;\delta(p)+(1-f)\;\delta(p-1)
  \;.
\end{align}

To see the evolution of the distribution and the vanishing of the SAC,
we compute the average and moments of $p$. It is useful to change
variables to $s_i{(\ell)}\equiv 2\,p_i{(\ell+1)}-1$, for which the
recursion Eq.~\eqref{eq:recursion_pi} reads
\begin{align}
  s_i{(\ell+1)}
  =
  {\widetilde C}^{g_i}_{100}\;s_{i_0}{(\ell)}
  +
  {\widetilde C}^{g_i}_{010}\;s_{i_1}{(\ell)}
  +
  {\widetilde C}^{g_i}_{001}\;s_{i_2}{(\ell)}
  +
  \cdots
  +
  {\widetilde C}^{g_i}_{111}\;s_{i_0}{(\ell)}\;s_{i_1}{(\ell)}\;s_{i_2}{(\ell)}
  \;,
\end{align}
with
\begin{align}
{\widetilde C}^{g_i}_{a}\equiv \frac{1}{2^3}\sum_{c=0}^7 (-1)^{a\cdot c}\;C^{g_i}_{c}
  \;,
\end{align}
where $a\cdot c\equiv a_0\,c_0+a_1\,c_1+a_2\,c_2$.

These relations allow us to compute the evolution of the moments
$\overline{s^q{(\ell)}}$. (Even if the distributions for the $s_i$ are
identical, independent of $i$, we keep some of the explicit indices
for bookkeeping of contractions.) The average
\begin{align}
  \overline{s{(\ell+1)}}
  &=
  \sum_{a=1}^7
  \overline{{\widetilde C}^{g_i}_{a}}\;
  \overline{[s_{i_0}{(\ell)}]^{a_0}}\,
  \overline{[s_{i_1}{(\ell)}]^{a_1}}\,
  \overline{[s_{i_2}{(\ell)}]^{a_2}}
    \nonumber\\
  &=
  \sum_{a=1}^7
  \overline{{\widetilde C}^{g_i}_{a}}\;
  \left[\overline{s{(\ell)}}\right]^{a_0+a_1+a_2}\,
  \;.
\end{align}
The recursion relating $\overline{s}(\ell+1)$ to $\overline{s}(\ell)$
depends on the gate set used for layer $\ell$ through the coefficients
$\overline{{\widetilde C}^{g_i}_{a}}$, which we present explicitly
below for the cases of inflationary and super-nonlinear gates.

Similarly, we compute the second moment
\begin{align}
  \overline{s^2{(\ell+1)}}
  =
  \sum_{a,b=1}^7
  \overline{{\widetilde C}^{g_i}_{a}\;{\widetilde C}^{g_i}_{b}}\;
  \overline{[s_{i_0}{(\ell)}]^{a_0+b_0}}\,
  \overline{[s_{i_1}{(\ell)}]^{a_1+b_1}}\,
  \overline{[s_{i_2}{(\ell)}]^{a_2+b_2}}
  \;.
\end{align}

We next consider explicitly the two classes of gates -- inflationary
and super-nonlinear gates -- that are deployed in the three-stage
cipher. For notational simplicity, we define the variables
$s(\ell)\equiv \overline{s(\ell)}$ and
$q(\ell)\equiv \overline{s^2(\ell)}$.

\subsubsection{Inflationary layers}
Upon computing the averages $\overline{{\widetilde C}^{g_i}_{a}}$ and
$\overline{{\widetilde C}^{g_i}_{a}\;{\widetilde C}^{g_i}_{b}}$ over
the 144 inflationary gates, the recursion relations read
\begin{subequations}
\begin{align}
  \label{eq:s-recursion}
  s(\ell+1)
  = \frac{2}{3}\,[s(\ell)]^2 + \frac{1}{3}\,[s(\ell)]^3
  \;,
\end{align}
\begin{align}
  q(\ell+1)
  =
  \frac{2}{3}\,[q(\ell)]^2 + \frac{1}{3}\,[q(\ell)]^3
  \;.
  \label{eq:recursion_inflationary_q}
\end{align}
\end{subequations}
The inflationary gates are special in that: (a) the equations for the
averages and second moments decouple; and more importantly (b) there
is no linear term in $q(\ell)$ in the equation for the second moment.

Note that the bimodal initial condition Eq.~\eqref{eq:initial-P},
where $p$ only takes values $p=0,1$, implies that an initial $q=1$
cannot evolve under Eq.~\eqref{eq:recursion_inflationary_q}, which
displays fixed points at $q=0,1$ (and a non-physical one at $q=-3$).
Hence, the value of $q$ cannot decay under the sole application of
inflationary gates. However, once nonlinear gates are applied and $q$
drops below 1, action by inflationary gates can accelerate its decay
to zero because of the absence of the linear term in $q(\ell)$ in the
recursion Eq.~\eqref{eq:recursion_inflationary_q}.

\subsubsection{Super nonlinear layers}
Using the averages $\overline{{\widetilde C}^{g_i}_{a}}$ and
$\overline{{\widetilde C}^{g_i}_{a}\;{\widetilde C}^{g_i}_{b}}$
computed over the 10752 super-nonlinear gates, the corresponding
recursion relations read
\begin{subequations}
\begin{align}
  s(\ell+1)
  =
  \frac{3}{7}\,s(\ell)
  +
  \frac{3}{7}\,[s(\ell)]^2
  +
  \frac{1}{7}\,[s(\ell)]^3
  \;,
\end{align}
\begin{align}
  \label{eq:recursion_supernonlinear_q}
  q(\ell+1)
  =&\;
  \frac{3}{28}\,
     \left([s(\ell)]^2 + [s(\ell)]^3\right)
     + \frac{3}{28}\,q(\ell) \,
     \left(1+ 2\,s(\ell) + 2\,[s(\ell)]^2\right)\\
   &+ \frac{3}{28}\,[q(\ell)]^2\,
     \left(1 + s(\ell)\right)
     + \frac{1}{28}\,[q(\ell)]^3
  \;.\nonumber
\end{align}
By contrast to the case of the inflationary gates, the recursion
relation for $q(\ell+1)$ above depends on both $s(\ell)$ and
$q(\ell)$, and, more importantly, it does contain a term linear in
$q(\ell)$. Note that, if the cipher only involved super-nonlinear
gates, the decay of $q$ would be at most exponential in $\ell$ because
of the linear terms in $q(\ell)$. In this case, reaching values of $q$
that are exponentially small in $n$ would require ${\cal O}(n)$
layers. By contrast, as we show below, the cipher with two
inflationary and one super-nonlinear stage reaches values of $q$ that
are exponentially small in $n$ with no more than ${\cal O}(\log n)$
layers of gates for each stage.

\end{subequations}
\end{widetext}

\subsubsection{Decay of SAC OTOC for the three-stage cipher}

We proceed sequentially stage-by-stage.
\\

\begin{enumerate}

\item{First inflationary stage:}

  As shown in Sec.~\ref{sec:inflation-with-tree}, after
  $\ell_{L_l}=\log_2 n$ layers of gates an initial string grows to
  macroscopic weight, corresponding to a fraction $f\to 1/2$ in the
  initial condition for the recursions for the two subsequent stages,
  or equivalently, $s\to 0$.~\footnote{One can arrive at a similar
    scaling for the number of layers $\ell_{L_l}$, but with a
    different prefactor, by deploying the recursion
    Eq.~\eqref{eq:s-recursion} with an initial condition $s(0)=1-2/n$,
    corresponding to a single flipped input. However, the recursion
    with this initial condition is not accurate because one cannot
    justify the assumption of independence of the bitline index. We
    therefore prefer to use the calculation of
    Sec.~\ref{sec:inflation-with-tree} to determine the depth
    $\ell_{L_l}$ needed to take $s\to 0$.}

\item{Super-nonlinear stage:}

  Taking $s\to 0$ simplifies the recursion relation for the second
  stage to:
\begin{align}
  \label{eq:recursion_supernonlinear_layers}
  q(\ell+1)
  =\;
  \frac{3}{28}\,q(\ell)
  + \frac{3}{28}\,[q(\ell)]^2\,
  + \frac{1}{28}\,[q(\ell)]^3
  \;,\hspace{-.6cm}
\end{align}
with $\ell$ in the range
$\ell_{L_l} \le \ell\le \ell_{L_l}+\ell_{N} -1$, and an initial
condition $q(\ell_{L_l})=1$.

\item{Final inflationary stage:}

  For the final stage, the recursion switches to:
\begin{align}
  \label{eq:recursion_inflationary_layers}
  q(\ell+1)
  =\;
  \frac{2}{3}\,[q(\ell)]^2\,
  + \frac{1}{3}\,[q(\ell)]^3
  \;,\hspace{-.6cm}
\end{align}
with $\ell$ in the range
$\ell_{L_l} + \ell_{N} \le \ell\le \ell_{L_l}+ \ell_{N} +
\ell_{L_r}-1$. The condition used to initiate this last stage,
$q(\ell_{L_l} + \ell_{N})$, is the final result of the previous stage.

\end{enumerate}

We can now examine the decay of the SAC OTOC from the final
$q(\ell_{L_l} + \ell_{N} + \ell_{L_r})$ resulting from iterating these
recursion relations. We shall place a bound on
$q(\ell_{L_l} + \ell_{N} + \ell_{L_r})$ as follows.

First, by using the initial condition $q(\ell_{L_l})=1$ in
Eq.~\eqref{eq:recursion_supernonlinear_layers}, we obtain
$q(\ell_{L_l}+1)=1/4$, which allows us to replace the recursion
Eq.~\eqref{eq:recursion_supernonlinear_layers}, for the subsequent
steps with $\ell$ in the range
$\ell_{L_l} +1 \le \ell\le \ell_{L_l}+\ell_{N} -1$, with the
inequality
\begin{align}
  q(\ell+1)
  &\le
  \frac{3}{28}\,q(\ell)
  + \frac{3}{28}\times \frac{1}{4}\,q(\ell)\,
  + \frac{1}{28}\times \frac{1}{16}\,q(\ell)
  \nonumber\\
  &=
  \frac{61}{448}\,q(\ell)
  \;.
\end{align}
Thus, after $\ell_{N}$ layers of super-nonlinear gates we obtain
$q(\ell_{L_l}+\ell_{N})\le
\frac{1}{4}\left(\frac{61}{448}\right)^{\ell_{N}-1}$. This value is
fed into the recursion for $q$ for the final inflationary stage,
Eq.~\eqref{eq:recursion_inflationary_layers}, which can in turn be
replaced by the inequality
\begin{align}
  q(\ell+1)
  \le \;
  [q(\ell)]^2\,
  \;,
\end{align}
to be used in the range
$\ell_{L_l} + \ell_{N} \le \ell\le \ell_{L_l}+ \ell_{N} +
\ell_{L_r}-1$.

We thus arrive at the upper bound for the final value of $q$, after
all $\ell_{L_l}+\ell_{N}+\ell_{L_r}$ layers of the three stages of the
cipher, namely
\begin{align}
  q(\ell_{L_l}+\ell_{N}+\ell_{L_r})\le
  \left[\frac{1}{4}\left(\frac{61}{448}\right)^{\ell_{N}-1}\right]^{2^{\ell_{L_r}}}
  \;.
  \label{eq:q_level}
\end{align}

We remark that the expression above was derived by using a continuum
mean-field approach, enabled by the tree structure of the cipher,
which should break down once $q$ drops below $2^{-n}$ -- at that
point, the action of the cipher can no longer be distinguished from
that of a random permutation via the SAC OTOC, even if one has access
to all the $2^n$ input/output relations. It follows from
Eq.~\eqref{eq:q_level} and our arguments for the depth of the first
inflationary layer above that $\ell_{L_l}=\ell_{L_r}=\log_2 n$ are
sufficient to comfortably achieve this criterium. (Recall that
$\ell_{N}=\log_3 n$ super-nonlinear layers are needed to saturate the
entropy.)

\begin{figure}[htp]
    \centering
      \includegraphics[angle=0,scale=.330]{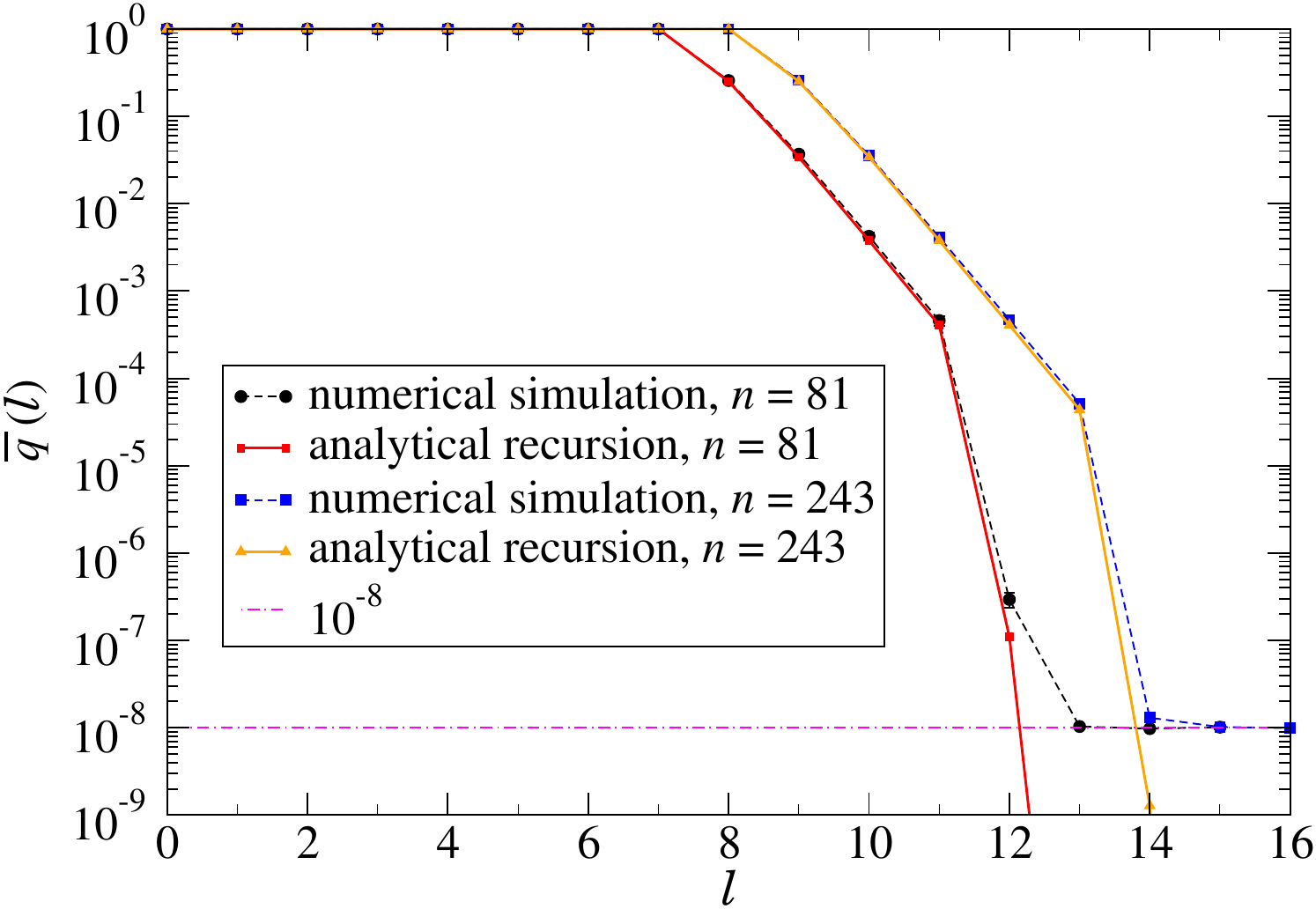}
      \caption{Square of the SAC OTOC, computed as a function of the
        number of applied layers, $\ell$, for three-stage ciphers with
        $\ell_{L_l}=\ell_{L_r}=\log_2 n$ and $\ell_{N}=\log_3 n$. The
        simulation is carried out for circuits with $n=81$ and
        $n=243$, and involves: (i) averaging over 56 ($n=81$) and 32
        ($n=243$) random and independent circuit realizations of the
        three-stage cipher; and (ii) sampling over $M=10^{8}$ input
        states selected randomly from the $2^{n}$ possible inputs. The
        numerical data points (dashed lines are guides to the eye) are
        compared to the predictions from the analytical recursion
        relations (shown as solid lines). The numerical results can
        only capture the behavior down to the floor level of
        $M^{-1}=10^{-8}$ (indicated by the horizontal dash-dotted
        line). Note that, consistent with analytical predictions,
        $q(\ell)$: (i) remains constant, equal to 1, throughout the
        first stage of inflation; (ii) decays approximately
        exponentially in $\ell$ after the first super-nonlinear layer;
        and (iii) drops faster than exponential in $\ell$ as soon as
        the last linear stage becomes active.  Also notice that the
        double-exponential decay of $q$ with $\ell$, predicted for
        this last stage, is difficult to capture with numerical
        simulation, and equally difficult to ascertain by an adversary
        with polynomial resources.  }
  \label{fig:OTOC_5I4S4I_infl}
\end{figure}

The predictions of the recursions derived above are in excellent
agreement with numerical simulation results shown in
Fig.~\ref{fig:OTOC_5I4S4I_infl}, which give us confidence in the
intuition we have built on the basis of our analytical approach about
the mechanisms responsible for the exponential (in $n$) decay of the
OTOCs. This intuition is summarized by the inequality
Eq.~\eqref{eq:q_level}. Most notably, removing the last stage of
inflationary gates would yield a $q$ that decays exponentially with
the number of layers, which only leads to polynomial in $n$ decay with
${\cal O}(\log n)$ layers of nonlinear gates, and would instead
require ${\cal O}(n)$ layers for exponential decay. The conclusion
would be unchanged had we removed both inflationary stages, in which
case the cipher would consist solely of nonlinear gates. (The decay of
the $q$ in this case is dominated by the tails in the stay-probability
of small strings discussed above.)  Also note that a single layer of
super-nonlinear gates, when flanked by the two inflationary stages
with ${\cal O}(\log n)$ layers, would be sufficient to suppress $q$
exponentially in $n$. However, in this case the entropy (generated
only by nonlinear gates), while extensive in $n$, would not saturate
the equilibrium value, which we argued requires at least
${\cal O}(\log n)$ layers of nonlinear gates. This last example
highlights that an OTOC can vanish exponentially while the output
function is still learnable in polynomial time, as discussed in
Sec.~\ref{sec:proliferation-with-tree}. In final analysis, this
supports our conclusion that the criterium for pseudo-randomness must
be that, up to an exponentially small fraction of atypical
(exponentially hard to identify) cases, OTOCs must vanish, and the
string entropies must saturate to their equilibrium values, including
their non-extensive contributions.

\section{Conclusions and open directions}
\label{sec:conclusions}

In this paper we presented a framework for analysis and design of
classical ciphers based on a close conceptual connection to the
problem of scrambling of quantum states. The close parallels to
quantum problems emerge naturally when we translate the actions of
measuring or flipping bits describing cipher attacks to Pauli string
operators. It is in string space that quantum mechanical-like
concepts, such as superposition of states and operator spreading,
become manifest. Here we studied the evolution of a classical
computation in string space and connected measures of cipher security,
such as resistance to plaintext and ciphertext attacks, to the
vanishing of out-of-time-order correlators (OTOCs), which in quantum
systems characterizes the appearance of chaos. Starting from any
initial string state, in both classical and quantum random circuits,
scrambling and chaos follow from the delocalization of amplitudes
across the exponentially large Hilbert space of string states. We
quantified this delocalization in string space in terms of
participation ratios and associated entropies. We argued that, in
order for a cipher to be indistinguishable from a random permutation
via polynomial number of queries, the OTOCs must vanish
super-polynomially in the number of bits and string entropies must
saturate. An important conclusion of this paper is that, for random
tree-structured three-stage ciphers built out of 3-bit gates in $S_8$,
OTOCs vanish exponentially and string entropies saturate for circuits
with as few as ${\cal O}(n\ln n)$ gates.

In order to reach this result we had to circumvent the effect of rare
events and the associated tails in the distribution of string weights,
connected with the slow growth of initially small strings. The same
problem appears in the analysis of scrambling by random quantum
circuits, in which case overcoming these tails requires larger
circuits. In particular, Brown and
Fawzi~\cite{Brown2013scrambling,Brown2015} prove scrambling and
decoupling for circuits with ${\cal O}(n\ln^2 n)$ gates, and Harrow
and Mehraban~\cite{Harrow2018approximate} prove that circuits laid
down in $D\sim \ln n$ dimensional lattices would scramble with
${\cal O}(n\ln n \ln\ln n)$ gates. (Ref.~\cite{Harrow2018approximate}
also conjectures that one should be able to achieve scrambling with
${\cal O}(n\ln n)$ gates, as we argued here.) The insight that allows
us to establish our result for the shorter circuit is to separate two
distinct processes controlling the evolution of string wave function
amplitudes, which we apply sequentially in a three-stage
structure. The first and the third circuits of the structure involve
the extension of small weight strings in the initial and final states
to strings of macroscopic weight (of order $3n/4$), processes
implemented via a set of 144 special linear gates (out of the $8!$
gates in $S_8$). These ``inflationary'' gates have the property that
the stay-probability for weight 1 substrings vanishes, ensuring that
strings gain weight at a sufficiently fast rate that the tails decay
exponentially with $n$ already for a circuit with ${\cal O}(n\ln n)$
gates. The second process, implemented by the middle layer of the
cipher (bookended by the two inflationary stages) involves applying
``super-nonlinear'' gates to heavy strings emerging from the
inflationary periods. During this string ``proliferation'' stage, the
application of a nonlinear gate splits a string state into a
superposition of multiple heavy string states, rapidly leading to an
exponential number of non-zero string amplitudes. As discussed in the
text, it is this proliferation in string space that leads to the
maximum entropies defined from the string state probabilities and to
the vanishing of OTOCs characterizing scrambling.

We reiterate that the diagnostics for the quality of a classical
cipher advanced in this paper are (i) the vanishing of OTOCs which
describe arbitrary (higher-order) differential attacks; and (ii) the
saturation of the string entropy. Even if an adversary cannot compute
the entropy for a specific permutation with polynomial resources (as
the dimension of string space is exponential), our statistical
mechanics approach allowed us to argue that the saturation of the
$n$-independent ${\cal O} (1)$ deficit in Eq.~\eqref{eq:equil-S}
requires at least ${\cal O} (\log n)$ layers of super-nonlinear
gates. As already mentioned in
subsection~\ref{sec:proliferation-with-tree}, this implies that the
nonlinear Boolean functions expressing the outputs of our three-stage
cipher following the action of $\ell _N\sim {\cal O} (\log n)$ layers
of super-nonlinear gates have degrees of up to $2^{\ell _N}$. There
are ${\cal O}(2^{2^{\ell_N}})$ coefficients in these polynomials,
making it impossible to learn these functions via fitting with
polynomial resources, another consistency argument for the security of
our cipher. This should be contrasted with the situation of a cipher
with a ${\cal O} (1)$ layers of nonlinear gates, which already
saturates the extensive part of the entropy (but not its non-extensive
correction), and generates polynomials of finite order that can be
learned with polynomial resources.

We end with comments on three issues that may inspire future
work. First, the three-stage structure that we employed in the
context of the classical cipher can also be implemented in a
tree-structured long-range quantum circuit, in which case we
expect that reaching the random Haar measure can also be realized in a
circuit with ${\cal O}(n\ln n)$ gates. It is already clear that this
can be achieved if one uses the same 144 linear 3-bit (or 3-qubit)
inflationary gates used above. We note that such a circuit would
scramble in depth (or time) ${\cal O}(\ln n)$, matching the scrambling
rate of a black
hole~\cite{Hayden2007,Sekino2008,Shenker2014,Shenker2014b,Maldacena2016,Stanford2016}. The
remaining question is whether there is a random unstructured circuit
of 2-qubit gates that can reach the same goal.

The second comment refers to the issue of averaging over circuits,
which also connects to discussions of $t$-design
(see~\cite{Harrow2018approximate,Roberts2017} and references
therein). It is most instructive to consider an explicit example,
namely that of the OTOC presented in Eq.~\eqref{eq:SAC}, which
measures the avalanche effect. For a given linear circuit [implemented
by a linear permutation,
$P_j(x\oplus 2^i)=P_j(x)\oplus P_j(2^i)\oplus b_j$, where $b_j=0,1$],
$C^{ij}_{\rm SAC}=(-1)^{P_j(2^i)\oplus b_j}=\pm 1$. For a {\it single}
realization of a random circuit, the $\pm 1$ value of the correlator
signals a weak cipher. By contrast, as discussed above, a good cipher
must be built out of nonlinear gates, in which case the correlator
$C^{ij}_{\rm SAC}$ vanishes for a given (sufficiently large)
circuit. It becomes immediately clear that if one averages over
circuits at this stage, the difference between the linear and
nonlinear circuits is washed out. This points to the dangers of
averaging too early, an issue well known from the theory of spin
glasses~\cite{FischerHertz1993,EdwardsAnderson}. Learning from the
latter, one should consider the behavior of a \textit{typical} circuit
by first squaring the correlator and only then averaging. This
suggests an Edwards-Anderson-like order parameter,
$q^{ij}_{\rm SAC} = \overline{\left(C^{ij}_{\rm SAC}\right)^2}$, which
is equal to 1 for linear circuits and vanishes for good ciphers. One
can repeat this discussion in the context of quantum circuits, where
averaging too early washes out the difference between Clifford and
universal unitary gates. The vanishing of the average of the correlator
$C^{ij}_{\rm SAC}$ is used to qualify Clifford circuits as 2-designs,
a designation that is physically meaningless as it does not help one
understand the behavior of a {\it typical} circuit. The difference
between Clifford and universal quantum circuits becomes apparent if
one uses the appropriate Edwards-Anderson-like order parameter, which
in the language of $t$-designs requires correlators with $t\ge 4$.

A related point, which follows from the discussions of delocalization
of wave function amplitudes in string space, is that once the system
is delocalized as signaled by the inverse participation ratio in
Eq.~\eqref{eq:P2}, it is delocalized by any other measure:
``Delocalized systems are all alike; every localized system is
localized in its own way.'' This suggests that once a system is a
$4$-design, it is also a $t$-design for $t>4$.

The third and final comment is that one can extend the discussion of
classical attacks expressed as OTOCs to quantum
attacks~\cite{quantum-attacks}. In the latter, the trace involved in
the computation of correlation functions (OTOCs) is replaced by
projective measurements following the action of certain operations
that can be translated into out-of-time-order operators (OTOOs), which
we plan to explore in a future publication.

In closing, fast ciphers with circuits of depth ${\cal O}(\ln n)$ are
intellectually interesting in their own right, as they saturate the
speed limit of information
scrambling~\cite{Hayden2007,Sekino2008,Shenker2014,Shenker2014b,Maldacena2016,Stanford2016}. Of
greater practical importance, however, is that they enable a
polynomial-overhead framework -- Encrypted-Operator Encryption -- for
secure computation directly on encrypted data, which we introduce in
Ref.~\cite{EOC} as an alternative to Homomorphic
Encryption~\cite{HE_PNAS2015}.

\begin{acknowledgments}
  The authors would like to thank Shiyu Zhou and Luowen Qian for
  useful discussions at the early stages of this paper, and Ran
  Canetti for many enlightening conversations and for stimulating us
  to explore physics-inspired frameworks for quantifying scrambling by
  classical ciphers. We also acknowledge David Huse and Zhi-Cheng Yang
  for insightful discussions.
  
\end{acknowledgments}

\appendix

\counterwithin{figure}{section}
\begin{widetext}

\section{Derivation of the OTOC in Eq.~\eqref{eq:CPCA} that is associated to a CPCA on a 3-round Feistel cipher}

Here we translate the attack presented by Patarin~\cite{Patarin} on a
3-round Feistel cipher [see Sec.~\ref{sec:feistel}, in particular
Eqs.~\eqref{eq:feistel-P} and \eqref{eq:feistel-Pinv}, for notation]
into an OTOC. The attack requires 3 oracle queries, 2 queries to the
encryption oracle, $P_{}$, and 1 query to the decryption oracle,
$P^{-1}_{}$:
\begin{enumerate}
\item Choose an input $\left(L_0^{(1)},R_0^{(1)}\right)$ and ask the
  encryption oracle for the output $\left(L_3^{(1)},R_3^{(1)}\right)=
  P_{}\left(L_0^{(1)},R_0^{(1)}\right)$;

\item Choose $L_0^{(2)}\ne L_0^{(1)}$ but reuse the same $R_0^{(1)}$
  value, and ask the encryption oracle for the output
  $\left(L_3^{(2)},R_3^{(2)}\right)=
  P_{}\left(L_0^{(2)},R_0^{(1)}\right)$;

\item Ask the decryption oracle for $\left(L_0^{(3)},R_0^{(3)}\right)=
  P^{-1}_{}\left(L_3^{(2)},R_3^{(2)}\oplus L_0^{(1)} \oplus L_0^{(2)} \right)$.
\end{enumerate}
For $r=3$, one can show using
Eqs.~(\ref{eq:feistel-P}, \ref{eq:feistel-Pinv}) that $R_0^{(3)} \oplus
R_0^{(1)} \oplus L_3^{(2)} \oplus L_3^{(1)} = 0$. 

To construct an OTOC that expresses the above constraint, it suffices
to use one bit, $i$, of the $R$ register and one bit, $j$, of the $L$
register. The correlation
\begin{align}
  \label{eq:CPCA-appendix}
  C^{ij}_{\rm CPCA}
  &=
  \tr\left[
    \rho_\infty\;
    \sx_j(0)\;\sx_i(\tau)
    \;\sz_i(0)\;\sx_i(\tau)
    \;\sz_j(\tau)\;\sx_j(0)
    \;\sz_j(\tau)\;\sz_i(0)
    \;
    \right]
  \nonumber\\
  &=
  (-1)^{\delta_{ij}}\;
  \tr\left[
    \rho_\infty
    \;\left(\sz_i(0)\;\sx_i(\tau)\right)^2
    \;\left(\sz_j(\tau)\;\sx_j(0)\right)^2
    \;
    \right]
  \;.
\end{align}
equals 1 if the evolution of the bit state is carried out via $P_{}$
and $P^{-1}_{}$. The correspondence between the attack and the OTOC is
extracted by following the sequence of operators from right-to-left in
the first line of Eq.~(\ref{eq:CPCA-appendix}):
\begin{enumerate}

\item $\sz_i(0)$ measures bit $i$ of $x_i$, or equivalently bit
  $i-n/2$ of $R_0^{(1)}$ ($R$ occupies the second half of the $n$-bit
  long block), through $(-1)^{x_i}$, at the initial input state,
  $\ket{x}=\ket{(L_0^{(1)}, R_0^{(1)})}$; 

\item $\sz_j(\tau)$ measures bit $j$ of $L_3^{(1)}$, through
  $(-1)^{P_j(x)}$, at the output and returns the system to the initial
  input state, $\ket{x}$; 

\item $\sx_j(0)$ flips the $j$-th bit of the initial input state,
  $\ket{x}$, into $\ket{x\oplus 2^j}$, thus setting $L_0^{(2)} =
  L_0^{(1)}\oplus 2^j$;

\item $\sx_i(\tau)\;\sz_j(\tau)$ first measures bit $j$ of
  $L_3^{(2)}$, $(-1)^{P_j(x\oplus 2^j)}$, at the output, then flips
  the $i$-th of the output state, thus shifting $R_3^{(2)}\to
  R_3^{(2)}\oplus 2^{i-n/2}$, and subsequently evolves the
  system backwards (via the $P^{-1}$) to the state
  $\ket{(L_0^{(3)},R_0^{(3)})}$;

\item $\sz_i(0)$ measures bit $i-n/2$ of $R_0^{(3)}$; and finally,

\item the product $\sx_j(0)\;\sx_i(\tau)$ undoes the flips above to
  restore the state to $\ket{x}$, so as to express the result as a
  trace over all possible inputs.
  
\end{enumerate}  

The correlator Eq.~\eqref{eq:CPCA-appendix} thus returns the average
over inputs of $ (-1)^{[R_0^{(3)}]_{i-n/2} + [R_0^{(1)}]_{i-n/2} +
  [L_3^{(2)}]_{j} + [L_3^{(1)}]_j} $; given the condition that
$R_0^{(3)} \oplus R_0^{(1)} \oplus L_3^{(2)} \oplus L_3^{(1)} = 0$ for
a 3-round Feistel cipher, it follows that
\begin{align}
  C^{i=j+n/2\;\;j}_{\rm CPCA} =1
  \;.
\end{align}
We have therefore successfully translated the attack on the 3-round
Feistel cipher in Ref.~\cite{Patarin} into an OTOC.

\section{Inflationary gates}
\label{sec:app:inflationary}

In the table below we list all the permutations in $S_8$ that are
associated to the 144 inflationary gates. These 3-bit gates are all
linear, and can be can be expressed in a circuit equivalent with 2-bit
CNOT gates as exemplified in Fig.~\ref{fig:CNOT-equiv-inflationary}.

\begin{table*}[h]
\renewcommand\thetable{OF INFLATIONARY GATES} \centering
\begin{tabular}{cc cc cc cc cc cc}
\hline
1:&\;0\,3\,5\,6\,7\,4\,2\,1\,\;\;\vline&
25:&\;1\,4\,6\,3\,7\,2\,0\,5\,\;\;\vline&
49:&\;2\,5\,7\,0\,1\,6\,4\,3\,\;\;\vline&
73:&\;4\,1\,2\,7\,3\,6\,5\,0\,\;\;\vline&
97:&\;5\,2\,3\,4\,0\,7\,6\,1\,\;\;\vline&
121:&\;6\,3\,1\,4\,5\,0\,2\,7\,\;\;\\ 2:&\;0\,3\,6\,5\,7\,4\,1\,2\,\;\;\vline&
26:&\;1\,4\,7\,2\,6\,3\,0\,5\,\;\;\vline&
50:&\;2\,5\,7\,0\,4\,3\,1\,6\,\;\;\vline&
74:&\;4\,1\,3\,6\,2\,7\,5\,0\,\;\;\vline&
98:&\;5\,2\,3\,4\,6\,1\,0\,7\,\;\;\vline&
122:&\;6\,3\,5\,0\,1\,4\,2\,7\,\;\;\\ 3:&\;0\,3\,7\,4\,5\,6\,2\,1\,\;\;\vline&
27:&\;1\,6\,2\,5\,4\,3\,7\,0\,\;\;\vline&
51:&\;2\,7\,1\,4\,5\,0\,6\,3\,\;\;\vline&
75:&\;4\,1\,3\,6\,7\,2\,0\,5\,\;\;\vline&
99:&\;5\,2\,6\,1\,0\,7\,3\,4\,\;\;\vline&
123:&\;6\,5\,0\,3\,1\,2\,7\,4\,\;\;\\ 4:&\;0\,3\,7\,4\,6\,5\,1\,2\,\;\;\vline&
28:&\;1\,6\,2\,5\,7\,0\,4\,3\,\;\;\vline&
52:&\;2\,7\,4\,1\,5\,0\,3\,6\,\;\;\vline&
76:&\;4\,1\,7\,2\,3\,6\,0\,5\,\;\;\vline&
100:&\;5\,2\,6\,1\,3\,4\,0\,7\,\;\;\vline&
124:&\;6\,5\,1\,2\,0\,3\,7\,4\,\;\;\\ 5:&\;0\,5\,3\,6\,7\,2\,4\,1\,\;\;\vline&
29:&\;1\,6\,4\,3\,2\,5\,7\,0\,\;\;\vline&
53:&\;2\,7\,5\,0\,1\,4\,6\,3\,\;\;\vline&
77:&\;4\,2\,1\,7\,3\,5\,6\,0\,\;\;\vline&
101:&\;5\,3\,0\,6\,2\,4\,7\,1\,\;\;\vline&
125:&\;6\,5\,1\,2\,3\,0\,4\,7\,\;\;\\ 6:&\;0\,5\,6\,3\,7\,2\,1\,4\,\;\;\vline&
30:&\;1\,6\,4\,3\,7\,0\,2\,5\,\;\;\vline&
54:&\;2\,7\,5\,0\,4\,1\,3\,6\,\;\;\vline&
78:&\;4\,2\,3\,5\,1\,7\,6\,0\,\;\;\vline&
102:&\;5\,3\,2\,4\,0\,6\,7\,1\,\;\;\vline&
126:&\;6\,5\,3\,0\,1\,2\,4\,7\,\;\;\\ 7:&\;0\,5\,7\,2\,3\,6\,4\,1\,\;\;\vline&
31:&\;1\,6\,7\,0\,2\,5\,4\,3\,\;\;\vline&
55:&\;3\,0\,4\,7\,5\,6\,2\,1\,\;\;\vline&
79:&\;4\,2\,3\,5\,7\,1\,0\,6\,\;\;\vline&
103:&\;5\,3\,2\,4\,6\,0\,1\,7\,\;\;\vline&
127:&\;7\,0\,1\,6\,2\,5\,4\,3\,\;\;\\ 8:&\;0\,5\,7\,2\,6\,3\,1\,4\,\;\;\vline&
32:&\;1\,6\,7\,0\,4\,3\,2\,5\,\;\;\vline&
56:&\;3\,0\,4\,7\,6\,5\,1\,2\,\;\;\vline&
80:&\;4\,2\,7\,1\,3\,5\,0\,6\,\;\;\vline&
104:&\;5\,3\,6\,0\,2\,4\,1\,7\,\;\;\vline&
128:&\;7\,0\,1\,6\,4\,3\,2\,5\,\;\;\\ 9:&\;0\,6\,3\,5\,7\,1\,4\,2\,\;\;\vline&
33:&\;1\,7\,2\,4\,6\,0\,5\,3\,\;\;\vline&
57:&\;3\,0\,5\,6\,4\,7\,2\,1\,\;\;\vline&
81:&\;4\,3\,1\,6\,2\,5\,7\,0\,\;\;\vline&
105:&\;5\,6\,0\,3\,2\,1\,7\,4\,\;\;\vline&
129:&\;7\,0\,2\,5\,1\,6\,4\,3\,\;\;\\ 10:&\;0\,6\,5\,3\,7\,1\,2\,4\,\;\;\vline&
34:&\;1\,7\,4\,2\,6\,0\,3\,5\,\;\;\vline&
58:&\;3\,0\,6\,5\,4\,7\,1\,2\,\;\;\vline&
82:&\;4\,3\,1\,6\,7\,0\,2\,5\,\;\;\vline&
106:&\;5\,6\,2\,1\,0\,3\,7\,4\,\;\;\vline&
130:&\;7\,0\,2\,5\,4\,3\,1\,6\,\;\;\\ 11:&\;0\,6\,7\,1\,3\,5\,4\,2\,\;\;\vline&
35:&\;1\,7\,6\,0\,2\,4\,5\,3\,\;\;\vline&
59:&\;3\,4\,0\,7\,5\,2\,6\,1\,\;\;\vline&
83:&\;4\,3\,2\,5\,1\,6\,7\,0\,\;\;\vline&
107:&\;5\,6\,2\,1\,3\,0\,4\,7\,\;\;\vline&
131:&\;7\,0\,4\,3\,1\,6\,2\,5\,\;\;\\ 12:&\;0\,6\,7\,1\,5\,3\,2\,4\,\;\;\vline&
36:&\;1\,7\,6\,0\,4\,2\,3\,5\,\;\;\vline&
60:&\;3\,4\,0\,7\,6\,1\,5\,2\,\;\;\vline&
84:&\;4\,3\,2\,5\,7\,0\,1\,6\,\;\;\vline&
108:&\;5\,6\,3\,0\,2\,1\,4\,7\,\;\;\vline&
132:&\;7\,0\,4\,3\,2\,5\,1\,6\,\;\;\\ 13:&\;0\,7\,3\,4\,5\,2\,6\,1\,\;\;\vline&
37:&\;2\,1\,4\,7\,5\,6\,3\,0\,\;\;\vline&
61:&\;3\,4\,5\,2\,0\,7\,6\,1\,\;\;\vline&
85:&\;4\,3\,7\,0\,1\,6\,2\,5\,\;\;\vline&
109:&\;6\,0\,1\,7\,3\,5\,4\,2\,\;\;\vline&
133:&\;7\,1\,0\,6\,2\,4\,5\,3\,\;\;\\ 14:&\;0\,7\,3\,4\,6\,1\,5\,2\,\;\;\vline&
38:&\;2\,1\,5\,6\,4\,7\,3\,0\,\;\;\vline&
62:&\;3\,4\,5\,2\,6\,1\,0\,7\,\;\;\vline&
86:&\;4\,3\,7\,0\,2\,5\,1\,6\,\;\;\vline&
110:&\;6\,0\,1\,7\,5\,3\,2\,4\,\;\;\vline&
134:&\;7\,1\,0\,6\,4\,2\,3\,5\,\;\;\\ 15:&\;0\,7\,5\,2\,3\,4\,6\,1\,\;\;\vline&
39:&\;2\,1\,5\,6\,7\,4\,0\,3\,\;\;\vline&
63:&\;3\,4\,6\,1\,0\,7\,5\,2\,\;\;\vline&
87:&\;4\,7\,1\,2\,3\,0\,6\,5\,\;\;\vline&
111:&\;6\,0\,3\,5\,1\,7\,4\,2\,\;\;\vline&
135:&\;7\,1\,2\,4\,0\,6\,5\,3\,\;\;\\ 16:&\;0\,7\,5\,2\,6\,1\,3\,4\,\;\;\vline&
40:&\;2\,1\,7\,4\,5\,6\,0\,3\,\;\;\vline&
64:&\;3\,4\,6\,1\,5\,2\,0\,7\,\;\;\vline&
88:&\;4\,7\,2\,1\,3\,0\,5\,6\,\;\;\vline&
112:&\;6\,0\,5\,3\,1\,7\,2\,4\,\;\;\vline&
136:&\;7\,1\,4\,2\,0\,6\,3\,5\,\;\;\\ 17:&\;0\,7\,6\,1\,3\,4\,5\,2\,\;\;\vline&
41:&\;2\,4\,1\,7\,5\,3\,6\,0\,\;\;\vline&
65:&\;3\,5\,0\,6\,4\,2\,7\,1\,\;\;\vline&
89:&\;4\,7\,3\,0\,1\,2\,6\,5\,\;\;\vline&
113:&\;6\,1\,0\,7\,3\,4\,5\,2\,\;\;\vline&
137:&\;7\,2\,0\,5\,1\,4\,6\,3\,\;\;\\ 18:&\;0\,7\,6\,1\,5\,2\,3\,4\,\;\;\vline&
42:&\;2\,4\,5\,3\,1\,7\,6\,0\,\;\;\vline&
66:&\;3\,5\,4\,2\,0\,6\,7\,1\,\;\;\vline&
90:&\;4\,7\,3\,0\,2\,1\,5\,6\,\;\;\vline&
114:&\;6\,1\,0\,7\,5\,2\,3\,4\,\;\;\vline&
138:&\;7\,2\,0\,5\,4\,1\,3\,6\,\;\;\\ 19:&\;1\,2\,4\,7\,6\,5\,3\,0\,\;\;\vline&
43:&\;2\,4\,5\,3\,7\,1\,0\,6\,\;\;\vline&
67:&\;3\,5\,4\,2\,6\,0\,1\,7\,\;\;\vline&
91:&\;5\,0\,2\,7\,3\,6\,4\,1\,\;\;\vline&
115:&\;6\,1\,3\,4\,0\,7\,5\,2\,\;\;\vline&
139:&\;7\,2\,1\,4\,0\,5\,6\,3\,\;\;\\ 20:&\;1\,2\,6\,5\,4\,7\,3\,0\,\;\;\vline&
44:&\;2\,4\,7\,1\,5\,3\,0\,6\,\;\;\vline&
68:&\;3\,5\,6\,0\,4\,2\,1\,7\,\;\;\vline&
92:&\;5\,0\,2\,7\,6\,3\,1\,4\,\;\;\vline&
116:&\;6\,1\,3\,4\,5\,2\,0\,7\,\;\;\vline&
140:&\;7\,2\,4\,1\,0\,5\,3\,6\,\;\;\\ 21:&\;1\,2\,6\,5\,7\,4\,0\,3\,\;\;\vline&
45:&\;2\,5\,1\,6\,4\,3\,7\,0\,\;\;\vline&
69:&\;3\,6\,0\,5\,4\,1\,7\,2\,\;\;\vline&
93:&\;5\,0\,3\,6\,2\,7\,4\,1\,\;\;\vline&
117:&\;6\,1\,5\,2\,0\,7\,3\,4\,\;\;\vline&
141:&\;7\,4\,0\,3\,1\,2\,6\,5\,\;\;\\ 22:&\;1\,2\,7\,4\,6\,5\,0\,3\,\;\;\vline&
46:&\;2\,5\,1\,6\,7\,0\,4\,3\,\;\;\vline&
70:&\;3\,6\,4\,1\,0\,5\,7\,2\,\;\;\vline&
94:&\;5\,0\,6\,3\,2\,7\,1\,4\,\;\;\vline&
118:&\;6\,1\,5\,2\,3\,4\,0\,7\,\;\;\vline&
142:&\;7\,4\,0\,3\,2\,1\,5\,6\,\;\;\\ 23:&\;1\,4\,2\,7\,6\,3\,5\,0\,\;\;\vline&
47:&\;2\,5\,4\,3\,1\,6\,7\,0\,\;\;\vline&
71:&\;3\,6\,4\,1\,5\,0\,2\,7\,\;\;\vline&
95:&\;5\,2\,0\,7\,3\,4\,6\,1\,\;\;\vline&
119:&\;6\,3\,0\,5\,1\,4\,7\,2\,\;\;\vline&
143:&\;7\,4\,1\,2\,0\,3\,6\,5\,\;\;\\ 24:&\;1\,4\,6\,3\,2\,7\,5\,0\,\;\;\vline&
48:&\;2\,5\,4\,3\,7\,0\,1\,6\,\;\;\vline&
72:&\;3\,6\,5\,0\,4\,1\,2\,7\,\;\;\vline&
96:&\;5\,2\,0\,7\,6\,1\,3\,4\,\;\;\vline&
120:&\;6\,3\,1\,4\,0\,5\,7\,2\,\;\;\vline&
144:&\;7\,4\,2\,1\,0\,3\,5\,6\,\;\;\\ 
\hline
\\
\end{tabular}
\caption{The lists of 8 numbers correspond to the outputs for each of
  the inputs 0 1 2 3 4 5 6 7, i.e., the corresponding permutation
  list. These 144 gates are found as follows. We compute the entries
  $t^{(g)}_{\alpha'\alpha}$ of the $64\times 64$ transition in
  Eq.~\eqref{eq:hopping} for each of the $8!$ gates in $S_8$. We
  filter out all gates for which the $9\times 9$ submatrix connecting
  the 9-dimensional subspace of weight 1 strings is non-zero, thus
  retaining only gates for which weight 1 substrings grow into weight
  2 or 3 substrings. All these gates are linear.}
\end{table*}

\begin{figure*}[h]
\includegraphics[width=0.4\textwidth]{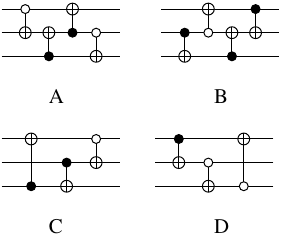}
\caption{Examples of 4 circuits written with 2-bit CNOT gates that are
  equivalent to the inflationary 3-bit gates. There are $3\times
  2^{4-1}=24$ circuits with the topology of the example shown in A,
  where the factor of 3 counts the number of ways of choosing the bit
  line with two controls, the factor of $2^4$ counts the choices of
  negated or not control (white or black control), and the factor of
  $2^{-1}$ accounts for the fact that exchanging the color on the
  controls of the two gates with the targets on the same bitline does
  not change functionality of the assembly of the four
  CNOTs. Similarly, there are $3\times 2^{4-1}=24$ circuits with the
  topology of B. Finally, there are $3!\times 2^3=48$ circuits similar
  to C and $3!\times 2^3=48$ similar to D. In total, there are 144
  such circuits.}
  \label{fig:CNOT-equiv-inflationary}
\end{figure*}

\end{widetext}

\bibliography{references}


\end{document}